\documentclass[aps,pra,twocolumn,superscriptaddress,nofootinbib]{revtex4-2}
\usepackage{hyperref}
\usepackage{graphicx}
\usepackage{physics}
\usepackage{mathrsfs}
\usepackage{amsmath}
\usepackage{amssymb}
\usepackage{amsfonts}
\usepackage{mathtools}
\usepackage{color}
\usepackage{lineno}
\hypersetup{colorlinks=true,citecolor=blue,linkcolor=blue,urlcolor=blue}

\newcommand{\subheading}[1]{\vspace{10pt}\noindent\textbf{#1}\\[2pt]}
\newcommand{\im}{{\rm i}}
\newcommand{\one}{{\rm I}}
\newcommand{\two}{{\rm I\hspace{-1.2pt}I}}

\begin{document}
\begingroup
\makeatletter
\let\addcontentsline\@gobblethree
\makeatother 
%\linenumbers
\title{Quantum Many-Body Mpemba Effect through Ruelle-Pollicott Resonances} 
\author{Shion Yamashika}
\affiliation{Department of Engineering Science, The University of Electro-Communications, Tokyo 182-8585, Japan.}
\author{Ryusuke Hamazaki}
\affiliation{Nonequilibrium Quantum Statistical Mechanics RIKEN Hakubi Research Team, RIKEN Pioneering Research Institute (PRI), 2-1 Hirosawa, Wako, Saitama 351-0198, Japan}
\affiliation{RIKEN Center for Interdisciplinary Theoretical and Mathematical Sciences (iTHEMS), 2-1 Hirosawa, Wako, Saitama 351-0198, Japan}

%Abstract – up to 200 words. 
%Main text – up to 3,000 words, excluding abstract, Methods, references and figure captions.
%Display items – up to 6 items (figures and/or tables). 
%Extended data – up to 10 items (figures and/or tables), appearing oncurve only.
%Figure captions should be fewer than 350 words, begin with a brief introductory sentence, and describe the meaning of all error bars.
%Articles can be divided into sections and subsections, but generic headings such as ‘Introduction’, ‘Results’ and ‘Discussion’ should be avoided. Headings must be fewer than 60 characters including spaces. 
%Concluding paragraphs that do no more than summarize the conclusions presented elsewhere in the manuscript are not permitted.
%Methods – up to 3,000 additional words, appearing online only. 
%References – up to 50.
%Articles include received/accepted dates. 
%Articles may be accompanied by supplementary information, which should be referred to explicitly in the main text. 

\begin{abstract}
Relaxation towards equilibrium is often assumed to be slower when a system starts farther from equilibrium, but this intuition fails in the Mpemba effect. Recent advances in controllable quantum platforms have enabled the exploration of its quantum analogue, the quantum Mpemba effect (QME), yet its microscopic origin remains largely unclear. Here we provide a general framework for understanding the QME in closed quantum many-body chaotic systems by reformulating the equilibration process of local subsystems in terms of Ruelle-Pollicott (RP) resonances. We show that suppressing the initial-state overlap with the dominant RP resonant mode accelerates subsystem equilibration and thereby yields the QME. We further uncover that a novel type of strong QME can occur via complete translation-symmetry breaking of initial states. We substantiate our predictions using the prototypical kicked Ising chain and a quantum automaton circuit. These findings cast the QME in closed many-body systems into a unified framework with open-system analogues and provide experimentally accessible signatures on state-of-the-art quantum platforms.
\end{abstract}
\maketitle

The Mpemba effect is an anomalous relaxation phenomenon, often summarised by the counterintuitive observation that hot water can freeze faster than cold water. Although reports of such phenomena date back to antiquity~\cite{Aristotle1923} and the effect was popularised in the modern literature after the rediscovery in the 1960s~\cite{Mpemba1969}, it remains a longstanding puzzle in nonequilibrium physics. Proposed explanations range from evaporative cooling~\cite{Vynnycky2010} to convection~\cite{Vynnycky2012,Vynnycky2015}, dissolved gases~\cite{Wojciechowski1988,Zimmerman2022}, and supercooling~\cite{Auerbach1995}. 
However, the universal origin of the Mpemba effect remains under debate due to its sensitivity to the experimental setup~\cite{Burridge2016,Burridge2020}.

The remarkable development of quantum platforms has renewed interest in the Mpemba effect in quantum regimes~\cite{Joshi2024,AharonyShapira2024,Zhang2025}. In quantum systems, the endpoint of relaxation can be specified as a well-defined target state (for instance, a thermal state), and proximity to it can be rigorously quantified using standard measures of distance between density matrices~\cite{Jozsa1994,Hbner1992,Fuchs1999}. This has led to the concept of the quantum Mpemba effect (QME), where, for the same dynamics, an initial state that is farther from the target state can nevertheless approach it faster than another initial state prepared closer. The QME not only provides a microscopic perspective on the anomalous relaxation, but also has practical relevance, for instance as a route to accelerated state preparation and cooling protocols in quantum platforms~\cite{Ares2025,Westhoff2025,Boubakour2025}.

\begin{figure}
\raggedright
\includegraphics[width=0.48\textwidth]{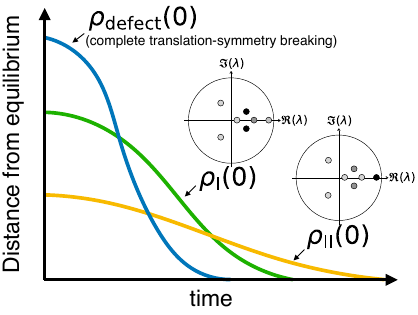}
\caption{\textbf{Quantum Mpemba effect through Ruelle-Pollicott resonances.} Late-time behaviour of a local subsystem in a closed quantum many-body chaotic system is controlled by the RP resonances, shown schematically as eigenvalues $\lambda$ in the complex plane (insets). The resonant mode closest to the unit circle decays most slowly and sets the bottleneck for subsystem equilibration. Therefore, suppressing its initial-state overlap, encoded in the colour intensity of the spectral points in the insets, accelerates local equilibration and yields the QME. In addition, when the translational symmetry is completely broken by the initial state, equilibration can be further accelerated, giving rise to a strong QME.}\label{fig:concept}
\end{figure}

The QME has been widely studied both in open~\cite{Carollo2021,Moroder2024,Nava2024,Chatterjee2023,Strachan2025,Chatterjee2024,Bao_2026,Caceffo_2024,Ares_2025} and closed~\cite{Ares2023,Turkeshi2025,Yamashika2024,Murciano2024,Liu2024,Yamashika2026,Yu_2025,Bhore_2025,Chalas_2024,Yamashika_2025,Tista2025,Klobas_2025,yzjd-pk8h,aresrylands2025} quantum systems. However, its microscopic origin remains elusive beyond a few special cases, such as integrable systems~\cite{Rylands2024} and Markovian open quantum systems~\cite{Carollo2021}. This gap is particularly acute for generic closed quantum systems exhibiting many-body chaos~\cite{Rigol2008,Eisert2015}, where the whole system evolves unitarily and never relaxes while a local subsystem relaxes to thermal equilibrium. Formulating a general description of such local equilibration is nontrivial, in contrast to the Markovian open quantum systems, where the equilibration process and the QME can be straightforwardly described by the Liouvillian spectrum~\cite{Gorini1976,Lindblad1976}.

Here, we fill this gap and propose a universal mechanism of the QME in closed quantum many-body chaotic systems, which can be understood in close analogy with its counterpart in Markovian open quantum systems. 
Our idea is to characterize local equilibration of the reduced density matrix in terms of Ruelle-Pollicott (RP) resonances, originally developed for describing correlation functions in classical chaos~\cite{Ruelle1986,Pollicott1985} and recently extended to quantum many-body systems~\cite{Prosen2002,nidari2024,Mori2024}. 
We uncover that the QME occurs when the initial state farther from equilibrium has a smaller overlap with slowly decaying RP resonant modes. Furthermore, the above general framework leads to an unconventional form of QME: strong QME via complete translation-symmetry breaking. That is, when the initial state completely breaks translational symmetry, the late-time relaxation law can be qualitatively modified, so which state relaxes faster is solely governed by this modified relaxation law. Such complete translation-symmetry breaking can be realized, for example, by inserting a local defect into the infinite temperature state. Figure~\ref{fig:concept} provides a visualisation of these results. We support our predictions with numerical simulations of the kicked Ising chain and of a quantum automaton circuit, which can be realised in state-of-the-art experimental platforms~\cite{Mi2021,Zhang2017,Seki2025,Levine2019,Evered2023}.

\subheading{Setup}
We consider a chain of $L$ qubits, whose state is described by the density matrix $\rho(t)=U^t \rho(0)U^{-t}$, where $\rho(0)$ is an initial state and $U$ is the unitary time-evolution operator. 
As schematically shown in Fig.~\ref{fig:setup}\textbf{a}, we decompose the whole system into a local subsystem $\mathrm{S}$ of finite length $\ell$ and its complement ${\rm E}$. The subsystem is described by the reduced density matrix 
\begin{align}\label{eq:rho_S(t)}
\rho_\mathrm{s}(t)=\Tr_{\rm E}[\rho(t)]=:\mathcal{U}_t \rho_\mathrm{s}(0). 
\end{align}
Although the whole system never relaxes due to the unitarity of its dynamics, under typical circumstances, the subsystem relaxes into an equilibrium state at large times in the thermodynamic limit as $\lim\limits_{t\to\infty} \lim\limits_{L\to\infty}\rho_\mathrm{s}(t)=\rho_\mathrm{s}^\mathrm{eq}$. In this context, the QME is said to occur when the subsystem that is initially farther from equilibrium relaxes faster than the one prepared closer to equilibrium. 

More precisely, the QME can be defined as follows. Let $\Delta_{\one}(t)$ and $\Delta_{\two}(t)$ denote the distances between the reduced density matrices at time $t$ and $\rho_{\rm s}^{\rm eq}$ for systems evolving under the same dynamics from two initial states labelled $\one$ and $\two$ (several choices of distance are possible; see Methods). If we assume that system $\one$ is initially farther from equilibrium, i.e., $\Delta_{\one}(0)>\Delta_{\two}(0)$, the QME is said to occur if and only if there exists a Mpemba time $t_{\rm M}>0$ such that~\cite{Ares2025}
\begin{equation}
\Delta_{\one}(t)<\Delta_{\two}(t)\quad\forall\,t>t_{\rm M}.
\label{eq:QME_late}
\end{equation}
Note that a reversal of the ordering of $\Delta_{\one}(t)$ and $\Delta_{\two}(t)$ at short times alone does not establish whether this condition is satisfied and hence whether the QME occurs, because the ordering may reverse again at later times. To determine when condition~\eqref{eq:QME_late} is satisfied, it is therefore necessary to examine the long-time asymptotic behaviour of the reduced density matrices.

If the reduced dynamics in equation~\eqref{eq:rho_S(t)} can be approximated as Markovian dynamics generated by the time-independent Lindblad equation,  the mechanism of the QME is understood from the spectral decomposition of the corresponding Liouvillian. That is, the initial state farther from equilibrium can relax faster simply because its overlap with the slowest decay mode is sufficiently small~\cite{Carollo2021,Zhang2025,Moroder2024} (see Methods). Separately, in integrable quantum many-body systems, the QME can be understood in terms of the charge transport properties of quasiparticles~\cite{Rylands2024}. However, in generic quantum many-body chaotic systems, such explanations fail because the reduced dynamics in equation~\eqref{eq:rho_S(t)} is typically non-Markovian and lacks the quasiparticle description. We here resolve this problem and show that the QME in closed quantum many-body chaotic systems can be understood within the framework analogous to that for Markovian dynamics, through the RP resonances. 

In what follows, we consider the global unitary dynamics $U$ that is invariant under $s$-site translations with $s\in \mathbb{N}$. We assume that the adjoint propagator $\mathcal{E}(\cdot)=U^\dagger(\cdot)U$ transforms a local operator to another local operator strictly within the light cone (see Fig.~\ref{fig:setup}\textbf{a}). We resolve it into quasi-momentum sectors $k$, $\mathcal{E}=\bigoplus_k \mathcal{E}_k$, where $\mathcal{E}_{k}$ is the restriction of $\mathcal{E}$ to the operator subspace spanned by Fourier-transformed operators $O_k:=\sum_{j\in \mathbb{Z}} e^{-\im kj}\mathcal{T}^{sj} O$, with $\mathcal{T}$ being the one-site translation map. 

RP resonances can be obtained by introducing a cutoff $r$ on operator size by defining a truncated propagator $\mathcal{E}_{k,r}=\Pi_{k,r} \mathcal{E}_k \Pi_{k,r}$ (see Fig.~\ref{fig:setup}\textbf{a})~\cite{Prosen2002,nidari2024}, where $\Pi_{k,r}$ denotes the projection onto the operator subspace spanned by $\{P_k|P\in \mathscr{P}_{r,s}\}$. Here $\mathscr{P}_{r,s}=\{\mathcal{T}^aP|P\in\mathscr{P}_r,\ a=0,1,...,s-1\}$ with $\mathscr{P}_r$ being the set of Pauli strings supported on at most $r$ consecutive qubits starting from the leftmost site of subsystem S; see Methods for their explicit expressions. Although $\mathcal{E}_k$ is strictly unitary, some eigenvalues of $\mathcal{E}_{k,r}$ do not approach the unit circle even when $r$ increases, as illustrated schematically in Fig.~\ref{fig:setup}\textbf{b}. The RP resonances are defined as those eigenvalues that remain strictly inside the unit circle even in the limit $r\to\infty$.

\begin{figure}
\includegraphics[width=0.5\textwidth]{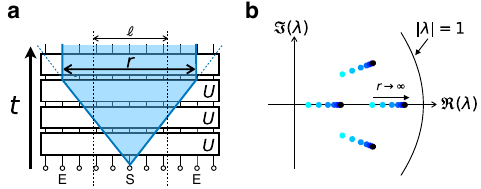}
\caption{\textbf{Truncated propagator and RP resonances.} \textbf{a,} Under the truncated propagator $\bigoplus_k \mathcal{E}_{k,r}$, any local observable increases its spatial extent strictly within the light cone until its support length reaches the cutoff $r$. \textbf{b,} As $r$ increases, the eigenvalues of $\mathcal{E}_{k,r}$ approach the unit circle because its original counterpart is unitary. The RP resonances are the eigenvalues that remain strictly inside the unit circle even in the limit $r\to\infty$. }\label{fig:setup}
\end{figure}

\subheading{QME through RP resonances}
Our central result is an asymptotic expansion of the reduced density matrix in terms of RP resonances. In the thermodynamic limit $L\to\infty$, we find for large $t$ that 
\begin{equation}
\rho_{\rm s}(t)\simeq \frac{I}{2^\ell}
+\sum_\alpha \int_{-\pi}^{\pi}\frac{dk}{2\pi}
\lambda_{\alpha,k}^{t} \Tr_{}[\rho(0)V_{\alpha,k}^{\mathrm{R}}]
\varrho_{\alpha,k}.
\label{eq:main_integral}
\end{equation}
Here $\lambda_{\alpha,k}$ and $V_{\alpha,k}^\mathrm{L(R)}$ are an eigenvalue and the corresponding left (right) eigenoperator of $\mathcal{E}_{k,r}$, with $\lambda_{\alpha,k}$ ordered in decreasing absolute value. The symbol $\simeq$ in equation~\eqref{eq:main_integral} reflects the finite cutoff $r$; the relation becomes exact in the limit $r\to\infty$. The operator $\varrho_{\alpha,k}$ acts only on the subsystem ${\rm S}$; see Methods for its explicit expression. Importantly, $\varrho_{\alpha,k}$ depends solely on 
$V_{\alpha,k}^\mathrm{L}$ and on the subsystem geometry, whereas all dependence on the initial condition enters through the overlap function $ \Tr_{}[\rho(0)V_{\alpha,k}^{\mathrm{R}}]$. The proof of equation~\eqref{eq:main_integral} is presented in Methods. 

Equation~\eqref{eq:main_integral} describes how a subsystem equilibrates as initially local correlations spread into its surroundings. RP resonances quantify the rate of this spreading through the decay rates of time correlation functions, which measure how long local correlation remains local~\cite{Prosen2002}. A resonant mode with a larger $|\lambda_{\alpha,k}|$ therefore corresponds to correlations that remain local for longer, and the overlap $\Tr_{}[\rho(0)V_{\alpha,k}^{\mathrm{R}}]$ with this mode measures how much of those correlations the initial state carries. Equation~\eqref{eq:main_integral} thus directly describes how the subsystem equilibrates as its initially local correlations spread into nonlocal ones.

When the initial state is invariant under $sl$-site translations, i.e., $\mathcal{T}^{sl} \rho(0)=\rho(0)$, the overlap function in equation~\eqref{eq:main_integral} develops delta-function peaks at the discrete momenta
$\Lambda_l=\{2\pi n/l\,|\,n=0,1,\dots,l-1\}$ as $ \Tr_{}[\rho(0)V_{\alpha,k}^{\mathrm{R}}]=2\pi c_{\alpha,k} \sum_{q\in \Lambda_l}\delta(k-q)$, where $c_{\alpha,k}$ are the weights of the resonant modes (fixed solely by $\rho(0)$; see Methods for the explicit expression). As a consequence, the $k$-integral in equation~\eqref{eq:main_integral} reduces to a finite sum and the reduced density matrix takes the form

\begin{equation}
\rho_\mathrm{s}(t)\simeq \rho_{\rm s}^{\rm eq}+\sum_{k\in\Lambda_l}\sum_{\alpha:\,|\lambda_{\alpha,k}|<1}
\lambda_{\alpha,k}^{t}\,c_{\alpha,k}\,\varrho_{\alpha,k}.
\label{eq:main_discrete}
\end{equation}
Here, the stationary part $\rho_{\rm s}^{\rm eq}$ is determined by the modes with $\lambda_{\alpha,k}=1$,
\begin{equation}
\rho_{\rm s}^{\rm eq}=\frac{I}{2^\ell}
+\sum_{k\in\Lambda_l}\sum_{\alpha:\,\lambda_{\alpha,k}=1}
c_{\alpha,k}\,\varrho_{\alpha,k}.
\label{eq:main_rhoeq}
\end{equation}

Equation~\eqref{eq:main_discrete} shows that the large-time behaviour of $\rho_\mathrm{s}(t)$ is governed by a finite set of effective decay modes associated with the RP resonances. 
This allows us to understand the QME in the closed quantum many-body chaotic system in direct analogy with the Markovian case. At large times, the subsystem dynamics is governed by the slowest decay mode, namely the RP resonance with the largest absolute value $|\lambda_{\alpha,k}|$. Consequently, a system that is initially farther from equilibrium can nevertheless relax faster if the weight of this slowest mode in the initial state, $|c_{\alpha,k}|$, is smaller. In other words, if we denote by $c_{\alpha,k}^{\one(\two)}$ the overlaps between the dominant resonant mode and the initial states labelled by $\one$ ($\two$), condition~\eqref{eq:QME_late} is satisfied if and only if $|c_{\alpha,k}^{\one}|<|c_{\alpha,k}^{\two}|$, yielding the necessary and sufficient conditions for the QME.

Given that the local correlations associated with the dominant RP resonant mode spread most slowly, the above condition requires the initial state closer to equilibrium to carry more of these local correlations. This explanation of the QME in terms of different initial weights of slow modes is common not only to the Markovian case but also to integrable systems~\cite{Rylands2024}.

A crucial difference from the Markovian case is that, in general, the weights $c_{\alpha,k}$ of the decay modes, and therefore the condition for the QME to occur, are determined not by the initial state of the subsystem of interest, $\rho_{\mathrm{s}}(0)$, but by that of the global system, $\rho(0)$. In addition, when the global dynamics $U$ has a conserved quantity $Q$, one has $\mathcal{E}Q=Q$, which implies the presence of an eigenvalue $\lambda_{\alpha,k}=1$~\cite{Duh2025} and therefore makes the equilibrium state~\eqref{eq:main_rhoeq} dependent on the initial state, even if the reduced dynamics in equation~\eqref{eq:rho_S(t)} has no conserved quantities. These features reflect the non-Markovian nature of the reduced dynamics in equation~\eqref{eq:rho_S(t)}. 

%We stress that our setup assumes neither weak system--environment coupling nor short environment memory. Thus, despite the formal similarity to the mode expansion for Markovian dynamics, the RP description does not generally imply that the reduced dynamics in equation~\eqref{eq:rho_S(t)} is Markovian.

In special cases, however, equation~\eqref{eq:main_integral} reduces to established frameworks, and, accordingly, so does the explanation of the QME in terms of RP resonances, as we demonstrate analytically in Supplementary Information Secs.~III and IV. For dual-unitary circuits, suitable initial preparations allow the reduced density matrix in equation~\eqref{eq:main_discrete} to be represented by repeated application of the same quantum channel, yielding Markovian dynamics (see also Methods). At the free-fermion point, we also show analytically how the spectral description underlying the RP resonances recovers the quasiparticle picture~\cite{Rylands2024}, a well-known description of the QME in integrable systems.

\subheading{QME in kicked Ising chain}
While the framework presented above generally holds whenever the unitary dynamics is invariant under $s$-site translations ($s\in \mathbb{N}$) and has the local structure, we now apply it to a concrete example. Specifically, we first focus on the chaotic kicked Ising chain with $s=1$, whose one-step dynamics is given by
\begin{align}\label{eq:KI}
U= e^{\im \tau \sum_{i} h_x X_i}
e^{\im \tau \sum_{i}(JZ_i Z_{i+1} + h_z Z_i)}, 
\end{align}
where $\{X_i,Y_i,Z_i\}$ are the Pauli matrices acting on the $i$-th qubit. 
For generic values of parameters $(\tau,h_x,h_z)$, the kicked Ising chain has no conserved quantity, which implies $|\lambda_{1,k}|<1$ for all $k$, and hence the subsystem relaxes into the infinite-temperature state $\rho_{\rm s}^{\rm eq}=I/2^\ell$~\cite{DAlessio2014,Fleckenstein2021}.

\begin{figure*}
\raggedright
\includegraphics[width=0.98\textwidth]{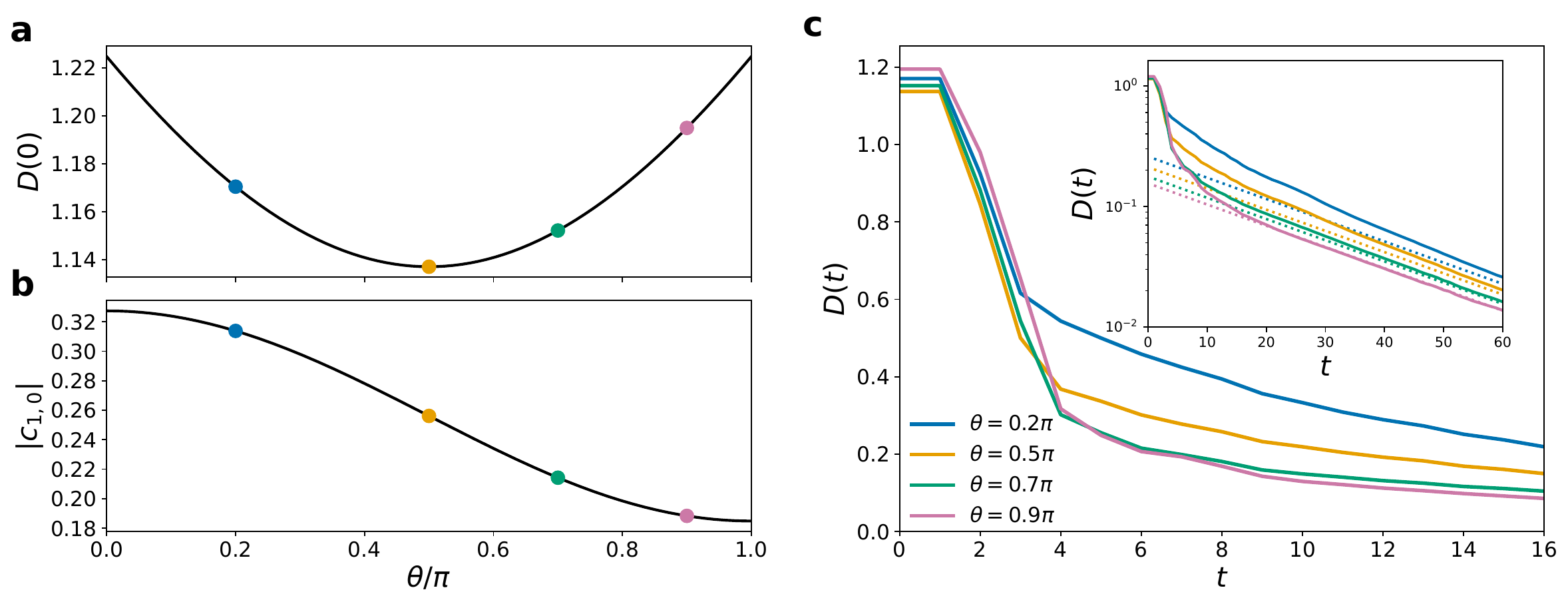}
\caption{\textbf{The QME in the dynamics of the kicked Ising chain.} \textbf{a,} The Bures distance $D(0)$ for the initial state~\eqref{eq:Psi_0}. It quantifies how far the system is from equilibrium at $t=0$. 
\textbf{b,} Weight $|c_{1,0}|$ of the slowest decay mode (the dominant RP resonance) for the same initial state. As $|c_{1,0}|$ decreases, the subsystem equilibrates faster. 
 \textbf{c,} Time evolution of the Bures distance under the dynamics of the kicked Ising chain. The solid curves correspond to the time evolution from the initial state in equation~\eqref{eq:Psi_0}. The inset shows the results for large times, in which the asymptotic forms of $D(t)$ predicted by equation~\eqref{eq:rho_S(t)_asymptotic} are plotted as the dotted curves. We set $(J,\tau,h_x,h_z)=(1,0.65,0.9,0.8)$, subsystem size $\ell=4$, and system size $L=24$ with periodic boundary conditions for all the plots. 
To obtain the solid curve in \textbf{b} and the dotted curves in the inset of \textbf{c}, we evaluate $\lambda_{1,k}$, $c_{1,k}$, and $\Tr_{}[\rho(0)V_{1,k}^\mathrm{R}]$ by numerically diagonalising $\mathcal{E}_{k,r}$ with $r=12$ (see Methods). Supplementary Information Secs.~II\,B, II\,A, and II\,C show, respectively, that changing $r$ barely changes the results, that finite-size effects are negligible over the plotted times, and that equation~\eqref{eq:rho_S(t)_asymptotic} also holds at other parameter values. The values of $\theta$ used in \textbf{c} are indicated by the filled circles in \textbf{a} and \textbf{b}. 
 }\label{fig:D_c_fidelity}  
\end{figure*}

The RP resonances of the kicked Ising chain~\eqref{eq:KI} have been studied in Ref.~\cite{nidari2024}, where it was shown that $|\lambda_{1,k}|$ attains its global maximum at $k=0$ and $\lambda_{1,0}\in \mathbb{R}$. It then follows from equation~\eqref{eq:main_discrete} that the reduced density matrix for the subsystem behaves at large times as 
\begin{align}\label{eq:rho_S(t)_asymptotic}
\rho_\mathrm{s}(t)\simeq \frac{I}{2^\ell} + \lambda_{1,0}^tc_{1,0} \varrho_{1,0}. 
\end{align}
This expression shows that the QME occurs when the global initial state of the system that is initially closer to equilibrium has a larger weight $|c_{1,0}|$ of the dominant RP resonance at $k=0$. 

To make the above prediction more concrete, we consider a family of initial states of the form $\rho(0)=\ket{\theta}\bra{\theta}$, where 
\begin{align}\label{eq:Psi_0}
\ket{\theta} = \cos(\frac{\theta}{2}) \ket{00...0}+  \sin(\frac{\theta}{2})  \ket{11...1}, 
\end{align}
with $\theta\in[0,\pi]$. We quantify the distance to equilibrium using the Bures distance~\cite{Hbner1992},
\begin{align}\label{eq:D(t)}
D(t):=\sqrt{2\qty(1-\Tr\sqrt{\sqrt{\rho_{\rm s}^{\rm eq}}\rho_{\rm s}(t)\sqrt{\rho_{\rm s}^{\rm eq}}})}. 
\end{align}
It quantifies how distinguishable two density matrices are (see Methods). 

On the one hand, as shown in Fig.~\ref{fig:D_c_fidelity}\textbf{a}, $D(0)$ for the initial state~\eqref{eq:Psi_0} attains its minimum at $\theta=\pi/2$ and the subsystem is initially closer to equilibrium when $\theta$ lies closer to $\pi/2$. On the other hand, Fig.~\ref{fig:D_c_fidelity}\textbf{b} shows the $\theta$-dependence of $|c_{1,0}|$, which is computed by numerically diagonalising $\mathcal{E}_{k,r}$ (see Methods), for the same family of initial states. We find that $|c_{1,0}|$ is maximal at $\theta=0$ and decreases as $\theta$ increases, indicating faster equilibration as $\theta$ approaches $\pi$. Taken together, these trends imply that the QME occurs for a pair of initial states with angles $\theta_i$ ($i=\one,\two$) if and only if the following two conditions are satisfied: (i) $|\theta_\one-\pi/2|<|\theta_\two-\pi/2|$ and (ii) $\theta_\one<\theta_\two$. 

We test these results in Fig.~\ref{fig:D_c_fidelity}\textbf{c}, in which we plot by solid curves the time evolution of $D(t)$ for the initial states~\eqref{eq:Psi_0} with several values of $\theta$, for the kicked Ising chain of length $L=24$ with periodic boundary conditions. It shows that curves corresponding to different values of $\theta$ intersect at certain times, signaling the occurrence of the QME. As expected, such intersections are observed only for pairs of initial states that satisfy both conditions (i) and (ii). The inset further shows that the exact numerical results (solid curves) approach the asymptotic form of $D(t)$ calculated from equation~\eqref{eq:rho_S(t)_asymptotic} (dotted curves) at late times. 
The residual discrepancy between the solid and dotted curves for small $\theta$ arises because the times shown are not yet sufficiently long for the leading asymptotic form to apply.

\subheading{Strong QME via translational symmetry}
While equation~\eqref{eq:main_discrete} suggests that the subsystem relaxes exponentially toward equilibrium, similarly to the Markovian case, this description breaks down when the initial state $\rho(0)$ completely breaks translational symmetry, i.e., when $\mathcal{T}^{sl}\rho(0)\neq\rho(0)$ for any integer $l$. In the translationally invariant case, the overlap factor $\Tr_{}[\rho(0)V_{\alpha,k}^\mathrm{R}] $ exhibits delta-function peaks, so that the $k$-integral in equation~\eqref{eq:main_integral} reduces to a discrete sum. By contrast, when translational symmetry is absent, such delta-function peaks can disappear in $\Tr_{}[\rho(0)V_{\alpha,k}^\mathrm{R}] $ and hence the contributions from the RP resonances are distributed over a continuum of momenta (see Fig.~\ref{fig:strong_QME}\textbf{a}). In this case, the large-time behaviour of the reduced density matrix is obtained by evaluating the integral over $k$ in equation~\eqref{eq:main_integral} via the method of steepest descent~\cite{Bender1999}, yielding
\begin{align}\label{eq:rho_S_NTI}
\rho_\mathrm{s}(t) \simeq \frac{I}{2^\ell} 
+
\sum_{\alpha} \sum_{k\in \Sigma_\alpha} \frac{\lambda_{\alpha,k}^t \Tr_{}[\rho(0)V_{\alpha,k}^\mathrm{R}] \varrho_{\alpha,k}}{\sqrt{2\pi t \nu_{\alpha,k}''} },
\end{align}
where $\nu_{\alpha,k}=-\ln \lambda_{\alpha,k}$, $\cdot '=\partial_k \cdot$, and $\Sigma_{\alpha}$  is the set of points at which $\nu_{\alpha,k}'=0$. The derivation of equation~\eqref{eq:rho_S_NTI} is provided in Supplementary Information Sec.~I.

\begin{figure*}
\raggedright
\includegraphics[width=\textwidth]{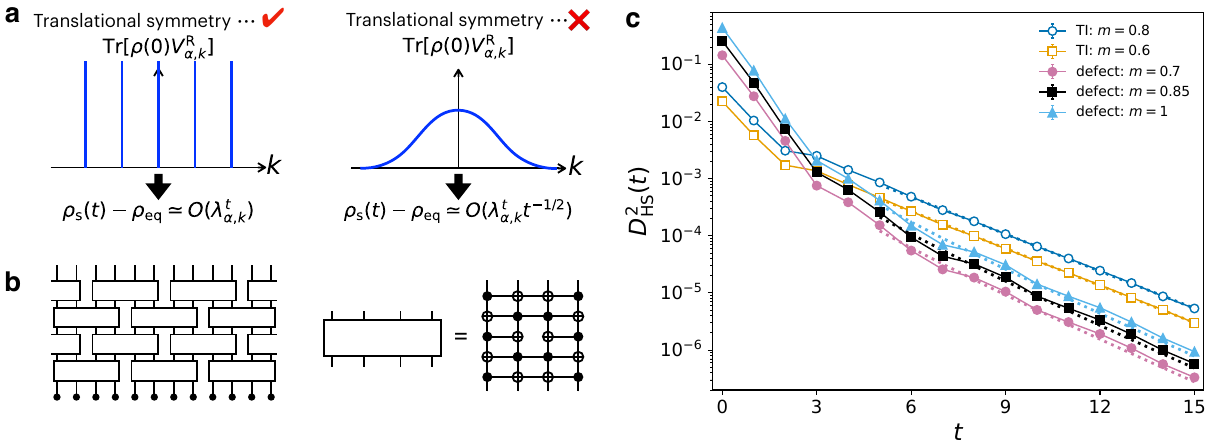}
\caption{\textbf{Strong QME from complete breaking of translational symmetry.}
\textbf{a,} Schematic contrast in the profile of the overlap function $\mathrm{Tr}[\rho(0)V^{\rm R}_{\alpha,k}]$. For an initial state with translational symmetry, the overlap develops discrete delta-function peaks at selected momenta, resulting in the purely exponential equilibration. By contrast, when translational symmetry is completely broken in the initial state, the overlap is instead spread over a continuum of momenta, which results in the accelerated equilibration with a factor $t^{-1/2}$. 
\textbf{b,} Schematic figure of the $s=4$ quantum automaton circuit (QAC) used to test the strong QME in Fig.~\ref{fig:strong_QME}. One Floquet step consists of two staggered layers of the same four-site gate consisting of CNOT and Toffoli gates.
\textbf{c,} Time evolution of $D_{\rm HS}^2(t)$ for the QAC starting from the initial state given by Eq.~\eqref{eq:rho_defect} (filled symbols), which decays faster than that for the translationally invariant state \eqref{eq:rho_TI} (empty symbols). The dotted curves correspond to the asymptotic form of $D_{\rm HS}^2(t)$ predicted by equations~\eqref{eq:rho_S(t)_asymptotic} and \eqref{eq:rho_S_NTI_KI} for the initial states $\rho_{\rm TI}$ and $\rho_{\rm defect}$, respectively. To obtain the dotted curves, we evaluate $\lambda_{1,k}$ and $\Tr[\rho(0)V_{1,k}^{\rm R}]$ by numerically diagonalising $\mathcal{E}_{k,r}$ with $r=19$. The Monte Carlo simulation uses $2^{27}$ samples for $\rho_{\rm TI}$ and $2^{33}$ samples for $\rho_{\rm defect}$. }\label{fig:strong_QME}
\end{figure*}

Equation~\eqref{eq:rho_S_NTI} highlights that complete translation-symmetry breaking can qualitatively reshape the relaxation law in two ways. First, the dominant resonant mode is no longer selected from the discrete momenta $\Lambda_l$ imposed by spatial periodicity of the initial state; instead it is determined by the set $\Sigma_\alpha$ of the points at which $|\lambda_{\alpha,k}|$ takes its local maxima. Second, $\rho_{\rm s}(t)$ acquires an algebraic prefactor $t^{-1/2}$ compared with the purely exponential form in equation~\eqref{eq:main_discrete}. These features imply that complete translational-symmetry breaking can induce a strong QME, in which the modified relaxation law determines which state equilibrates faster, rather than by a mere difference in their overlaps with the same slowest mode.

We note that equation~\eqref{eq:rho_S_NTI} has exceptions and limitations. In Supplementary Information Sec.~VII, we show that, for random product initial states, the overlap $\Tr_{}[\rho(0)V_{\alpha,k}^{\mathrm R}]$ is not smooth in $k$, so the steepest-descent method is not applicable. In this case, the relaxation of $\rho_{\rm s}(t)$ typically acquires a $t^{-1/4}$ algebraic prefactor. We also show in Supplementary Information Sec.~VI that, for initial states with a long but finite spatial period of $l$ unit cells ($l\gg1$), the algebraically corrected relaxation described by equation~\eqref{eq:rho_S_NTI} can appear at intermediate times, but ultimately crosses over to the purely exponential decay of equation~\eqref{eq:rho_S(t)_asymptotic}.

To make the above argument explicit, we consider the quantum automaton circuit (QAC) with $s=4$ illustrated in Fig.~\ref{fig:strong_QME}\textbf{b}, whose time evolution can be simulated directly in the thermodynamic limit by Monte Carlo sampling (see Supplementary Information Sec.~V\,A). As in the kicked Ising chain, $|\lambda_{1,k}|$ for this model attains its global maximum at $k=0$ and $\lambda_{1,k}\in\mathbb{R}$ in the vicinity of $k=0$. For initial states with the translation symmetry of the dynamics, the late-time reduced density matrix is therefore given by equation~\eqref{eq:rho_S(t)_asymptotic}. When this symmetry is completely broken, equation~\eqref{eq:rho_S_NTI} instead yields
\begin{align}\label{eq:rho_S_NTI_KI}
\rho_\mathrm{s}(t) \simeq \frac{I}{2^\ell}
+\frac{\lambda_{1,0}^t\Tr_{}[\rho(0)V^{\rm R}_{1,0}]\varrho_{1,0}}{\sqrt{2\pi t\nu_{1,0}''}}.
\end{align}
Comparing this expression with equation~\eqref{eq:rho_S(t)_asymptotic}, one readily finds that the relaxation in the QAC is algebraically accelerated when the initial state completely breaks translational symmetry.

For the numerical test of the above results, we consider the following two initial states:
\begin{align}
\rho_{\rm TI}&=\bigotimes_{i=0\,{\rm mod}\,4}\rho_Z(m)\bigotimes_{i\ne0\,{\rm mod}\,4}\frac{I}{2},\label{eq:rho_TI}\\
\rho_{\rm defect}&=\bigotimes_{i\in\mathcal D}\rho_Z(m)\bigotimes_{i\notin\mathcal D}\frac{I}{2},\label{eq:rho_defect}
\end{align}
where $\rho_Z(m)=(I+mZ)/2$ and $\mathcal D$ consists of three adjacent sites within the subsystem S. Figure~\ref{fig:strong_QME}\textbf{c} shows the Monte Carlo simulation for the Hilbert-Schmidt distance $D_{\rm HS}^2(t)=\Tr[(\rho_{\rm s}(t)-I/2^\ell)^2]$ from equilibrium. The numerical results agree well with the predictions obtained from equations~\eqref{eq:rho_S(t)_asymptotic} and~\eqref{eq:rho_S_NTI_KI}. As expected, the system starting from the initial state in equation~\eqref{eq:rho_defect} relaxes faster despite starting farther from equilibrium, indicating the occurrence of the strong QME rooted in the complete breaking of translational symmetry.

We stress that the strong QME in Fig.~\ref{fig:strong_QME}\textbf{c} arises from a mechanism fundamentally distinct from that of the QME in Fig.~\ref{fig:D_c_fidelity}.
Here, the overlap profile of $\rho_{\rm TI}$ has a delta-function peak at $k=0$, whereas that of $\rho_{\rm defect}$ is a continuous density with no such peak. Their relaxation therefore cannot be compared simply in terms of different weights of the same slowest mode, as in Fig.~\ref{fig:D_c_fidelity}. Moreover, the two states share the same exponential decay rate and hence equations~\eqref{eq:rho_S(t)_asymptotic} and~\eqref{eq:rho_S_NTI_KI} give $D_{{\rm HS},{\rm defect}}/D_{{\rm HS},{\rm TI}}=O(t^{-1/2})\to0$ as $t\to\infty$, making clear that the algebraic correction causes $\rho_{\rm defect}$ to relax faster at late times.

\subheading{Discussion}
To summarise, we have shown that the QME in closed quantum many-body chaotic systems can be understood within the framework based on RP resonances. These resonances govern the relaxation of local subsystems, and the QME occurs when the state initially farther from equilibrium carries a smaller weight of the dominant (slowest) resonant mode. We have also shown that a complete breaking of translational symmetry in the initial state can qualitatively reshape the late-time relaxation law and induce a strong QME: which initial state equilibrates faster is then dictated by this change in the asymptotic relaxation law, rather than by a mere difference in constant prefactors.

These results provide a way to predict the QME for a previously unstudied model or family of initial states from the relevant RP modes and initial-state properties, without computing their time evolution, as demonstrated here for the kicked Ising chain and the QAC. The kicked Ising chain has already been realised in a variety of experimental platforms, such as trapped ions and superconducting quantum processors~\cite{Mi2021,Zhang2017,Seki2025}. The QAC considered here can likewise be implemented using CNOT and CCNOT gates, both of which can be experimentally demonstrated on neutral-atom platforms~\cite{Levine2019,Evered2023}. An experimental test would directly probe the mechanism of local equilibration in quantum many-body chaotic systems through RP resonances.

More broadly, the present formalism provides a route to analyse late-time local relaxation beyond the QME by giving access to the reduced density matrix rather than a particular observable. In particular, it enables principled predictions for the late-time behaviour of nonlinear functionals of density matrices, such as quantum Fisher information, R\'enyi entanglement entropies, and other information-theoretic quantities, which are not captured by tracking any single observable.  It will also be interesting to extend our framework to continuous-time Hamiltonian dynamics, where one expects an analogous mode picture to govern late-time local equilibration, now shaped by energy conservation and the hydrodynamic slow modes. Another direction is to explore long-range interacting systems, where locality and the effective light cone are modified; understanding how the resonance picture adapts may reveal new late-time relaxation laws.
The QME in quantum chaotic systems has also been studied in Haar random circuits~\cite{Turkeshi2025,Liu2024} that lack the discrete translation symmetry assumed in the present formalism. How these studies connect to our framework is therefore a nontrivial open question. A connection to the membrane picture~\cite{Zhou2020} for entanglement dynamics of quantum chaotic systems is likewise nontrivial: our framework describes late-time relaxation at fixed subsystem size, whereas the membrane picture describes the leading behaviour when both subsystem size and time are large. Understanding how the membrane picture can be reconciled with the formalism of RP resonances is another interesting direction for future work.

\subheading{Methods}
\textbf{QME in Markovian open quantum systems.}
If the reduced dynamics in equation~\eqref{eq:rho_S(t)} can be approximated as Markovian evolution, so that it forms a semigroup $\mathcal{U}_t\mathcal{U}_{t'}=\mathcal{U}_{t+t'}$ ($t,t'>0$), the mechanism of the QME becomes transparent.
In this case $\rho_\mathrm{s}(t)$ obeys the Gorini-Kossakowski-Sudarshan-Lindblad (GKSL) master equation~\cite{Gorini1976,Lindblad1976},
\begin{align}
\dot{\rho}_\mathrm{s}(t)=-\mathcal{L}\rho_\mathrm{s}(t),
\end{align}
where $\mathcal{L}$ is a Liouvillian superoperator. Then, under natural assumptions such as uniqueness of the equilibrium state, $\rho_\mathrm{s}(t)$ can be written at large times as 
\begin{align}
\rho_\mathrm{s}(t)\simeq \rho_\mathrm{s}^{\rm eq} + e^{-\mu_\mathrm{sl} t }\rho_\mathrm{sl} \Tr_{}[ \rho_{\rm s}(0)\pi_\mathrm{sl}^\dag].
\end{align}
Here, $\mu_\mathrm{sl}$ is the eigenvalue of $\mathcal{L}$ with the smallest nonzero real part, which we here assume to be real for simplicity, and $\pi_\mathrm{sl}$ ($\rho_\mathrm{sl}$) denotes the corresponding left (right) eigenvector.  
As a consequence, a subsystem that is initially farther from equilibrium can nevertheless relax faster, for the simple reason that its overlap with the slowest decay mode, $\Tr_{}[\rho_{\rm s}(0)\pi_\mathrm{sl}^\dag ]$, is sufficiently small~\cite{Carollo2021,Zhang2025,Moroder2024}. 
\\
~
\\
\textbf{Asymptotic expression of density matrix.}
Here we provide the derivation of equations~\eqref{eq:main_integral} and \eqref{eq:main_discrete}. Without loss of generality, $\rho_{\rm s}(t)$ can be written as 
\begin{align}\label{eq:rho_s_TP}
\rho_{\rm s}(t)=\frac{1}{2^\ell}\left(I+\mkern-12mu\sum_{\substack{n\in\mathbb Z,\ P\in\mathscr P_{\ell,s}\\ \operatorname{supp}(\mathcal T^{sn}P)\subseteq\mathrm S}}\mkern-12mu\Tr[\rho(t)\mathcal T^{sn}P]\mathcal T^{sn}P\right),
\end{align}
where
\begin{align}
\mathscr P_{r,s}&=\{\mathcal T^aP\mid P\in\mathscr P_r,\ a=0,\ldots,s-1\},\\
\mathscr P_r&=\left\{\bigotimes_{j=0}^{r-1}P_j\ \middle|
\begin{array}{l}
P_0\in\{X,Y,Z\},\\
P_j\in\{I,X,Y,Z\},\\
1\leq j<r
\end{array}\right\}.
\end{align}
For example, $\mathscr P_{r,s}$ and $\mathscr P_r$ for small values of $r$ and $s$ specifically read
\begin{align}
\mathscr P_1&=\{X,Y,Z\},\\
\mathscr P_2&=\begin{aligned}[t]&\{XI,XX,XY,XZ,YI,YX,\\
&\quad YY,YZ,ZI,ZX,ZY,ZZ\},\end{aligned}\\
\mathscr P_{1,2}&=\{X,Y,Z,\mathcal T X,\mathcal T Y,\mathcal T Z\}.
\end{align}
The condition $\operatorname{supp}(\mathcal T^{sn}P)\subseteq\mathrm S$ in equation~\eqref{eq:rho_s_TP} means that all nonidentity factors of the translated Pauli string $\mathcal T^{sn}P$ lie inside the subsystem S. For $s=1$, this is equivalent to taking the summation of $\mathcal T^iP$ over $i=0,\ldots,\ell-1$ and $P\in\mathscr P_{\ell-i}$.

Using the Fourier transform $P_k=\sum_{n\in\mathbb Z}e^{-\im kn}\mathcal T^{sn}P$, the trace in equation~\eqref{eq:rho_s_TP} can be rewritten as
\begin{align}\label{eq:tr[rhoTP]}
\Tr_{}[\rho(t) \mathcal{T}^{sn} P] 
= 
\int_{-\pi}^\pi \frac{dk}{2\pi} e^{\im k n} \Tr_{}[\rho(t) P_k]. 
\end{align}
For large times, $\Tr_{}[\rho(t) P_k]$ can be expressed in terms of RP resonant modes as~\cite{Prosen2002,nidari2024} 
\begin{align}\label{eq:tr[rhoPk]}
\Tr_{}[\rho(t) P_k]\simeq \sum_\alpha \lambda_{\alpha,k}^t (v_{\alpha,P_k}^\mathrm{L})^* \Tr_{}[\rho(0)V^\mathrm{R}_{\alpha,k}]. 
\end{align}
Here, $\lambda_{\alpha,k}$ denote the eigenvalues of $\mathcal{E}_{k,r}$ and $V^{\rm L(R)}_{\alpha,k}=\sum_{P\in \mathscr{P}_{r,s}}v_{\alpha,P_k}^{\rm L(R)}P_k$ are the corresponding left (right) eigenoperators. They satisfy $\mathcal E_{k,r}V_{\alpha,k}^{\rm R}=\lambda_{\alpha,k}V_{\alpha,k}^{\rm R}$ and $\mathcal E_{k,r}^\dagger V_{\alpha,k}^{\rm L}=\lambda_{\alpha,k}^*V_{\alpha,k}^{\rm L}$. We order $\lambda_{\alpha,k}$ by decreasing absolute value and adopt the bi-orthonormal convention $\sum_{P\in\mathscr P_{r,s}}(v_{\alpha,P_k}^{\rm L})^*v_{\beta,P_k}^{\rm R}=\delta_{\alpha,\beta}$.
Substituting equations~\eqref{eq:tr[rhoTP]} and \eqref{eq:tr[rhoPk]} into equation~\eqref{eq:rho_s_TP}, we arrive at equation \eqref{eq:main_integral} with 
\begin{align}
\varrho_{\alpha,k}=\frac{1}{2^\ell}\sum_{\substack{n\in\mathbb Z,\ P\in\mathscr P_{\ell,s}\\ \operatorname{supp}(\mathcal T^{sn}P)\subseteq\mathrm S}}e^{\im kn}(v_{\alpha,P_k}^{\rm L})^*\mathcal T^{sn}P.
\end{align}

Equation~\eqref{eq:main_integral} further simplifies when $\mathcal{T}^{sl} \rho(0)=\rho(0)$ for some integer $l$. For such initial states, the overlap function $\Tr_{}[\rho(0) V_{\alpha,k}^\mathrm{R}]$ reduces to 
\begin{align}\label{eq:Tr_delta}
\Tr_{}[\rho(0) V_{\alpha,k}^\mathrm{R}]=2\pi c_{\alpha,k} \sum_{q\in \Lambda_l} \delta(k-q), 
\end{align}
where 
\begin{align}
c_{\alpha,k}=l^{-1}\sum_{n=0}^{l-1}\sum_{P\in\mathscr P_{r,s}}v_{\alpha,P_k}^{\rm R}\Tr[\rho(0)\mathcal T^{sn}P]e^{-\im kn}.
\end{align}
Substituting equation~\eqref{eq:Tr_delta} into equation~\eqref{eq:main_integral} and performing the integral over $k$, we obtain equation~\eqref{eq:main_discrete}.  
\\~\\
\textbf{Truncated propagator.} Here, we construct $\mathcal E_{k,r}$. Let us assume that $\mathcal E(\cdot)=U^\dagger(\cdot)U$ enlarges the support of a local operator by at most $\delta r$ sites on each side in one time step. For $P\in\mathscr P_{r,s}$, this locality gives
\begin{align}\label{eq:propagator_pauli_expansion}
\mathcal E(P)=\sum_{\substack{n\in\mathbb Z\\P'\in\mathscr P_{r+2\delta r,s}}}M_{P',P}^{(n)}\mathcal T^{sn}P'.
\end{align}
Here, $M_{P',P}^{(n)}$ is the coefficient of the output string $\mathcal T^{sn}P'$ in $\mathcal E(P)$. Equivalently, $M_{P',P}^{(n)}=2^{-|B|}\Tr_B[(\mathcal T^{sn}P')^\dagger U^\dagger P U]$, where $B$ is a finite region containing the supports of both $\mathcal E(P)$ and $\mathcal T^{sn}P'$. Since $\mathcal E(P)$ is local, $M_{P',P}^{(n)}=0$ for sufficiently large $|n|$.

Substituting $P_k=\sum_{m\in\mathbb Z}e^{-\im km}\mathcal T^{sm}P$ into $\mathcal E$, we find
\begin{align}
\mathcal E(P_k)=\sum_{m\in\mathbb Z}e^{-\im km}\mathcal E(\mathcal T^{sm}P).
\end{align}
Using $[\mathcal E,\mathcal T^s]=0$, it can be rewritten as
\begin{align}\label{eq:propagator_fourier_commutation}
\mathcal E(P_k)=\sum_{m\in\mathbb Z}e^{-\im km}\mathcal T^{sm}\mathcal E(P).
\end{align}
Substituting equation~\eqref{eq:propagator_pauli_expansion} into equation~\eqref{eq:propagator_fourier_commutation}, one finds that $\mathcal E$ preserves each quasi-momentum sector, i.e., $\mathcal E=\bigoplus_k\mathcal E_k$, where
\begin{align}\label{eq:propagator_momentum_sector}
\mathcal E_k(P_k)=\sum_{P'\in\mathscr P_{r+2\delta r,s}}M_{P',P}(k)P'_k,
\end{align}
with
\begin{align}
M_{P',P}(k)=\sum_{n\in\mathbb Z}e^{\im kn}M_{P',P}^{(n)}.
\end{align}
The truncated propagator within the operator subspace spanned by $\{P_k\mid P\in\mathscr P_{r,s}\}$ can be obtained by neglecting $P'\in\mathscr P_{r+2\delta r,s}\setminus\mathscr P_{r,s}$ in equation~\eqref{eq:propagator_momentum_sector}, yielding
\begin{align}\label{eq:E_kr}
\mathcal E_{k,r}(P_k)=\sum_{P'\in\mathscr P_{r,s}}M_{P',P}(k)P'_k.
\end{align}
Using equation~\eqref{eq:E_kr}, we extract the RP resonances and the corresponding eigenoperators by numerically diagonalising $M(k)$. This strategy is justified by the strict locality of $U$, which guarantees that the action of $\mathcal E_k$ is recovered from $\mathcal E_{k,r}$ once $r$ is taken sufficiently large. 
\\~\\
\textbf{Distances to equilibrium.}
For the analyses of the kicked Ising model, we employ the Bures distance~\eqref{eq:D(t)} to quantify the distance from equilibrium of the subsystem. The Bures distance has good properties as a measure of how close the subsystem is to the equilibrium state: it is non-negative $D(t)\geq0$ and vanishes if and only if $\rho_{\rm s}(t)=\rho_{\rm s}^{\rm eq}$.

For the kicked Ising chain, we can explicitly derive the asymptotic form of the Bures distance at large times by using the fact that the equilibrium state is $\rho_{\rm s}^{\rm eq}=I/2^\ell$. In this case, the right-hand side of equation~\eqref{eq:D(t)} can be expanded in terms of $\delta \rho_{\rm s}(t)= \rho_{\rm s}(t)-\rho^{\rm eq}_{\rm s}$ as 
\begin{align}\label{eq:D(t) large time}
D(t)\simeq 2^{\ell/2-1} \sqrt{\Tr_{} [\delta \rho_{\rm s}(t)^2]}. 
\end{align}
Substituting equation~\eqref{eq:rho_S(t)_asymptotic} into the above equation, we obtain 
\begin{align}\label{eq:D(t) large time KI TI}
D(t)\simeq 2^{\ell/2-1} \abs{c_{1,0}} \abs{\lambda_{1,0}}^t \sqrt{\Tr_{}[\varrho_{1,0}^2]}, 
\end{align}
which applies when $\rho(0)$ has translational symmetry or spatial periodicity with finite period (note that $\varrho_{1,0}$ is Hermitian for our model). On the other hand, when $\rho(0)$ has a smooth momentum-space overlap, the Bures distance at large times can be obtained by substituting equation~\eqref{eq:rho_S_NTI_KI} into equation~\eqref{eq:D(t) large time} as 
\begin{align}\label{eq:D(t) large time KI NTI}
D(t)\simeq \frac{2^{\ell/2-1}|\Tr_{}[\rho(0) V_{1,0}^{\rm R}]| \abs{\lambda_{1,0}}^t \sqrt{\Tr_{}[\varrho_{1,0}^2]}}{\sqrt{2\pi \nu_{1,0}'' t}}. 
\end{align} 
The dotted curves in the inset of Fig.~\ref{fig:D_c_fidelity}\textbf{c} are obtained from equation~\eqref{eq:D(t) large time KI TI}.

For the QAC, we use the squared Hilbert--Schmidt distance from $\rho_{\rm s}^{\rm eq}=I/2^\ell$,
\begin{equation}\label{eq:D_HS(t)}
D_{\rm HS}^2(t)=\Tr\bigl[(\rho_{\rm s}(t)-I/2^\ell)^2\bigr]=\Tr[\rho_{\rm s}(t)^2]-2^{-\ell}.
\end{equation}
This quantity also has good properties as a measure of distance to equilibrium, i.e., $D_{\rm HS}^2\geq0$ and $D_{\rm HS}^2=0$ if and only if $\rho_{\rm s}(t)=\rho_{\rm s}^{\rm eq}$. In addition, its expression in equation~\eqref{eq:D_HS(t)} allows us to compute the distance from the subsystem purity by Monte Carlo sampling directly in the thermodynamic limit (see Supplementary Information Sec.~V\,A). The predictions obtained from the RP resonances in Fig.~\ref{fig:strong_QME}\textbf{c} are obtained by substituting equations~\eqref{eq:rho_S(t)_asymptotic} and~\eqref{eq:rho_S_NTI_KI} into equation~\eqref{eq:D_HS(t)}.

\par\vspace{\baselineskip}\noindent
\textbf{RP resonances at the dual-unitary point.}
The kicked Ising model at $J=h_x=1$ and $\tau=\pi/4$ is known to be a dual-unitary circuit with $s=2$, in which the single-site reduced density matrix follows Markovian dynamics with suitable initial conditions~\cite{Bertini2026}. At this point, we can derive exactly the RP resonances contributing to one-site reduced dynamics (see Supplementary Information Sec.~IV) as
\begin{equation}\label{eq:DU_main_resonances}
\lambda_{\pm,k}=e^{\pm\im k}\cos^2\phi,
\end{equation}
where $\phi=2\tau h_z$. Their left and right eigenoperators can also be derived as
\begin{align}
V_{-,k}^{\rm L}&=X_{-,k}+\cdots,
\label{eq:DU_main_left_minus}\\
V_{+,k}^{\rm L}&=X_{+,k}-Y_{+,k}\tan\phi+\cdots,
\label{eq:DU_main_left_plus}\\
V_{-,k}^{\rm R}&=X_{-,k}+Y_{-,k}\tan\phi+\cdots,
\label{eq:DU_main_right_minus}\\
V_{+,k}^{\rm R}&=X_{+,k}+\cdots.
\label{eq:DU_main_right_plus}
\end{align}
Here $X_{\pm,k}$ and $Y_{\pm,k}$ are the Fourier-transformed Pauli operators on the odd ($+$) and even ($-$) sublattices, and the ellipses denote the terms whose support lengths are larger than 1 (see Supplementary Information Sec.~IV).

We consider the initial state
\begin{align}
\rho(0)&=\bigotimes_{j\in\mathbb Z}\left(\frac{I_{2j}}{2}\otimes\sigma_{2j+1}\right),
\label{eq:DU_main_initial}\\
\sigma_{2j+1}&=\sigma=\frac{I+xX+yY+zZ}{2}.
\label{eq:DU_main_sigma}
\end{align}
This initial state has a finite overlap only with $V_{+,0}^{\rm R}$, and the terms with support lengths larger than 1 in equation~\eqref{eq:DU_main_right_plus} do not contribute to the overlap $\Tr_{}[\rho(0)V_{+,k}^{\rm R}]$. As a result, substituting equations~\eqref{eq:DU_main_resonances}--\eqref{eq:DU_main_sigma} into equation~\eqref{eq:main_integral}, we obtain the exact one-site reduced dynamics for $t\geq1$,
\begin{equation}\label{eq:DU_main_rho}
\rho_{\rm s}(t)=\Phi_\phi^t(\sigma),
\end{equation}
with
\begin{equation}\label{eq:DU_main_channel}
\Phi_\phi(\bullet)=\frac{\Tr(\bullet)}{2}I+\frac{\Tr(\bullet X)}{2}\cos^2\phi\,\bigl(X-\tan\phi\,Y\bigr).
\end{equation}
Equation~\eqref{eq:DU_main_rho} shows that the reduced density matrix for a single site follows Markovian dynamics.

In the ordered operator basis $\{I,X,Y,Z\}$, the matrix representation of $\Phi_\phi$ is
\begin{equation}\label{eq:DU_main_channel_matrix}
[\Phi_\phi]=
\begin{pmatrix}
1&0&0&0\\
0&\cos^2\phi&0&0\\
0&-\cos\phi\sin\phi&0&0\\
0&0&0&0
\end{pmatrix}.
\end{equation}
Its eigenvalues are $1,\cos^2\phi,0,0$. Note that the eigenvalue $\cos^2\phi$, which determines the relaxation rate in equation~\eqref{eq:DU_main_rho}, coincides with the RP resonances in equation~\eqref{eq:DU_main_resonances} at $k=0$, selected by the two-site translation symmetry of the initial state in equation~\eqref{eq:DU_main_initial}. In addition, its left and right eigenoperators
\begin{equation}\label{eq:DU_main_channel_eigenoperators}
L_\phi=X,\qquad R_\phi=X-Y\tan\phi,
\end{equation}
also coincide with the one-site components of $V_{+,0}^{\rm R}$ and $V_{+,0}^{\rm L}$ in equations~\eqref{eq:DU_main_right_plus} and~\eqref{eq:DU_main_left_plus}, respectively. Note that the operator and density-matrix propagators are adjoints, and hence the right eigenoperator of $\Phi_\phi$ corresponds to the left eigenoperator of the RP resonance, and vice versa. Thus, in the present setting, the RP resonance and its mode directly reduce to the Liouvillian spectrum governing Markovian dynamics.

These results also yield an explicit condition for the QME. Using equation~\eqref{eq:DU_main_rho}, we find that the squared Hilbert--Schmidt distance from the equilibrium state $I/2$ is given by $D_{\rm HS}^2(0)=(x^2+y^2+z^2)/2$ and $D_{\rm HS}^2(t)=x^2|\cos\phi|^{4t-2}/2$ for $t\geq1$. Thus, the QME occurs if the initial state with the larger $x^2+y^2+z^2$ has the smaller $|x|$, which is the weight of the RP resonant mode $V_{+,k}^{\rm R}$ at $k=0$ in the global initial state or, equivalently, the overlap of the initial state of the subsystem with the Liouvillian eigenoperator $L_\phi$.

\vspace{10pt}\noindent\textbf{Data availability}\\*[2pt]
Source Data are provided with this paper. 

\subheading{Code availability}
The computer codes used to generate the results reported in this paper will be made publicly available. 

\vspace{10pt}\noindent\textbf{Acknowledgements}\\*[2pt]
S.\,Y. was supported by JSPS KAKENHI Grant No. JP25K23355 and by Institute for Advanced Science, University of Electro-Communications. R.\,H. was supported by JSPS KAKENHI Grant No. JP24K16982 and by JST ERATO Grant Number JPMJER2302, Japan. The authors used OpenAI Codex to assist with manuscript revision and take full responsibility for the final manuscript.

\subheading{Author Contributions}
S.Y. and R.H. contributed to developing the theory, interpreting the results, and writing the manuscript. S.Y. performed the numerical simulations and analytic computations.

\subheading{Competing interests}
The authors declare no competing interests.

\clearpage
\endgroup
\makeatletter
\@booleanfalse\twocolumn@sw
\onecolumngrid
\setcounter{NAT@ctr}{0}
\let\arxivlabel\label
\renewcommand{\label}[1]{\ifnum\pdfstrcmp{#1}{LastBibItem}=0 \arxivlabel{SM-LastBibItem}\else\arxivlabel{#1}\fi}
\renewcommand{\theHequation}{SM.\arabic{equation}}
\renewcommand{\theHfigure}{SM.\arabic{figure}}
\renewcommand{\theHtable}{SM.\arabic{table}}
\renewcommand{\theHsection}{SM.\arabic{section}}
\hypersetup{pageanchor=false}
\makeatother
 
\setcounter{page}{1}
\onecolumngrid
\newcounter{equationSM}
\newcounter{figureSM}
\newcounter{tableSM}
\stepcounter{equationSM}
\setcounter{equation}{0}
\setcounter{figure}{0}
\setcounter{table}{0}
\setcounter{section}{0}
\makeatletter
\renewcommand{\theequation}{\textsc{sm}-\arabic{equation}}
\renewcommand{\thefigure}{\textsc{sm}-\arabic{figure}}
\renewcommand{\thetable}{\textsc{sm}-\arabic{table}}

\begin{center}
  {\large{\bf Supplementary Information for \\~\\
  ``Quantum Many-Body Mpemba Effect through Ruelle-Pollicott Resonances''}}
\end{center} 
%\vspace{20pt}
\tableofcontents

\section{Asymptotic analysis of the reduced density matrix}\label{sec:Laplace}
We analyse the large-time behaviour of the reduced density matrix in the form 
\begin{align}
\rho_{\rm s}(t)-\frac{I}{2^\ell}=\sum_\alpha \int_{-\pi}^\pi \frac{dk}{2\pi} \lambda_{\alpha,k}^t \Tr[\rho(0) V_{\alpha,k}^\mathrm{R}] \varrho_{\alpha,k}. \label{eq:integral}
\end{align}
As argued in the main text, when $\rho(0)$ has a finite spatial periodicity, $\Tr_{}[\rho(0)V_{\alpha,k}^{\rm R}]$ reduces to the sum of delta functions, and equation~\eqref{eq:integral} becomes a discrete sum, leading to purely exponential relaxation. Here we focus on the opposite case in which $\rho(0)$ has no finite spatial periodicity, i.e., the translational symmetry is completely broken. 

It is convenient to rewrite $\lambda_{\alpha,k}$ as 
\begin{align}
\lambda_{\alpha,k} = e^{-\nu_{\alpha,k}}, 
\end{align}
Then, equation~\eqref{eq:integral} becomes 
\begin{align}
\rho_{\rm s}(t)-\frac{I}{2^\ell}=\sum_\alpha \int_{-\pi}^\pi \frac{dk}{2\pi} e^{-t\nu_{\alpha,k}} \Tr[\rho(0) V_{\alpha,k}^\mathrm{R}] \varrho_{\alpha,k}. \label{eq:integral 2}
\end{align}
We first consider the case in which the integrand in the above equation admits analytic continuation to $k\in \mathbb{C}$. We assume that the integration contour can be deformed without crossing singularities. Let $\Sigma_\alpha$ denote the nondegenerate saddle points at which $\nu_{\alpha,k}'=0$. When $\Tr_{}[\rho(0)V_{\alpha,k}^{\rm R}]\varrho_{\alpha,k}\neq0$ at the saddle points, by expanding $\nu_{\alpha,k}$ in the vicinity of $q\in\Sigma_\alpha$ as 
\begin{align}\label{eq:lambda_aq}
\nu_{\alpha,k}=\nu_{\alpha,q}+\frac{\nu_{\alpha,q}''}{2}(k-q)^2+O((k-q)^3), 
\end{align}
we can apply the steepest descent method~\cite{ref:BenderOrszag} to equation~\eqref{eq:integral 2}. It results in
\begin{align}\label{eq:asymptotic}
\rho_{\rm s}(t)-\frac{I}{2^\ell}
\simeq
\sum_\alpha\sum_{q\in\Sigma_\alpha}
\frac{\lambda_{\alpha,q}^t\Tr_{}[\rho(0)V_{\alpha,q}^{\rm R}]\varrho_{\alpha,q}}{\sqrt{2\pi t\nu_{\alpha,q}''}}.
\end{align}
This gives equation~\eqref{eq:rho_S_NTI} in the main text. 

When the prefactor $\Tr_{}[\rho(0)V_{\alpha,q}^{\rm R}]\varrho_{\alpha,q}$ vanishes at a saddle point $q\in\Sigma_\alpha$, the relaxation can acquire a different algebraic correction. Assuming that the prefactor is analytic near $q$, we expand it as
\begin{align}\label{eq:overlap_expansion}
\Tr[\rho(0)V_{\alpha,k}^{\rm R}]\varrho_{\alpha,k}
=\sum_{n\geq0}\eta_{\alpha,n}(k-q)^n,
\qquad \gamma_q=\min\{n:\eta_{\alpha,n}\neq0\}. 
\end{align}
On the one hand, if $\gamma_q$ is even, the leading term follows from the Gaussian integral. On the other hand, for odd $\gamma_q$, this term integrates to zero, and we retain the next term in the prefactor and the cubic term in $\nu_{\alpha,k}$~\cite{nemes2013}. The steepest descent method then gives
\begin{align}\label{eq:local_asymptotic}
\rho_{\rm s}(t)-\frac{I}{2^\ell}
\simeq
\sum_\alpha\sum_{q\in\Sigma_\alpha}
\lambda_{\alpha,q}^t
\times
\begin{dcases}
\frac{\Gamma((\gamma_q+1)/2)}{2\pi}
\qty(\frac{2}{t\nu_{\alpha,q}''})^{(\gamma_q+1)/2}
\eta_{\alpha,\gamma_q},
&\gamma_q=\text{even},
\\
\frac{\Gamma((\gamma_q+2)/2)}{2\pi}
\qty(\frac{2}{t\nu_{\alpha,q}''})^{(\gamma_q+2)/2}
\qty[
\eta_{\alpha,\gamma_q+1}
-
\frac{(\gamma_q+2)\nu_{\alpha,q}'''}{6\nu_{\alpha,q}''}
\eta_{\alpha,\gamma_q}],
&\gamma_q=\text{odd}.
\end{dcases}
\end{align}
Here $\Gamma(x)$ is the Gamma function. When the prefactor $\Tr_{}[\rho(0)V_{\alpha,k}^{\rm R}]\varrho_{\alpha,k}$ is not analytic at the saddle points, a different relaxation law can arise, as discussed in Sec.~\ref{sec:random_product}.

We note that numerically verifying the algebraic correction to the relaxation law obtained by the steepest descent method requires large systems and can therefore be computationally demanding. For a periodic chain of $sL$ sites with $s$-site translation symmetry, the allowed momenta are separated by $\Delta k=2\pi/L$. For a dominant real stationary point $q$, the Gaussian envelope has width $\delta k\sim[t\nu_{\alpha,q}'']^{-1/2}$. To resolve this envelope and approximate the momentum sum by an integral, one therefore needs 
\begin{align}
\frac{2\pi}{L}\ll \frac{1}{\sqrt{t\nu_{\alpha,q}''}},
\qquad\text{or equivalently}\qquad
L\gg 2\pi\sqrt{t\nu_{\alpha,q}''}.
\label{eq:momentum_resolution}
\end{align}
Thus, larger systems are required at later times, and this condition must remain satisfied in the long-time regime where the leading RP resonance becomes dominant. For the quantum automaton circuit (QAC) discussed in Sec.~\ref{sec:QCA}, this difficulty can be overcome and the algebraic correction in the relaxation law can be verified through Monte Carlo simulations in the thermodynamic limit.

\section{Numerical checks in the kicked Ising model}
\label{sec:KI_checks}

\subsection{Finite-size effects}
\label{sec:KI_finite_size}

\begin{figure}[h]
\centering
\includegraphics[width=\textwidth]{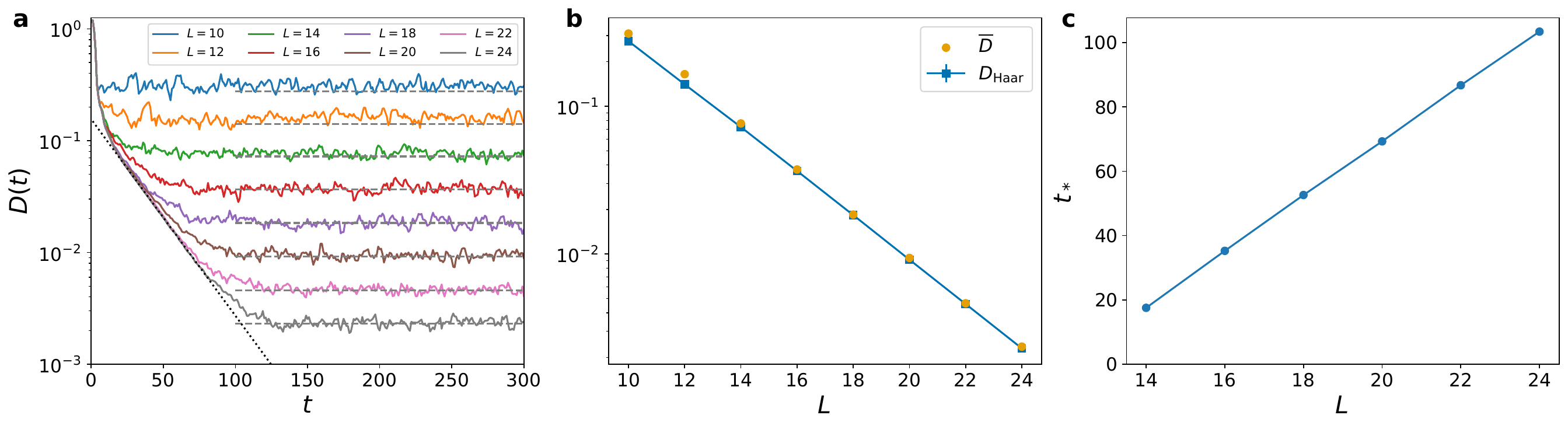}
\caption{\textbf{Finite-size effects in the kicked Ising chain.} \textbf{a,} Bures distance for $L=10,12,\ldots,24$. The black dotted curve shows $D_{\rm RP}(t)$ calculated from the dominant RP resonance and its mode at $r=12$. Grey dashed lines show $D_{\rm Haar}$. \textbf{b,} Time averages $\overline D=[t_{\rm max}-t_{\rm min}+1]^{-1}\sum_{t=t_{\rm min}}^{t_{\rm max}}D(t)$ with $t_{\rm min}=150$ and $t_{\rm max}=300$ (circles) and $D_{\rm Haar}$ obtained by averaging over Haar-random pure states in the translation-invariant and reflection-symmetric sector (squares). To calculate $D_{\rm Haar}$, we sample 5,000 states for each $L\leq20$ and 2,000 states for each $L=22,24$. The error bars of the finite sample average are smaller than the symbols. \textbf{c,} Time $t_*$ as a function of $L$, calculated from equation~\eqref{eq:KI_time_window}. We use $(J,\tau,h_x,h_z)=(1,0.65,0.9,0.8)$, $\ell=4$, and the initial state $\ket{\theta}=\cos(\theta/2)\ket{0}^{\otimes L}+\sin(\theta/2)\ket{1}^{\otimes L}$ with $\theta=0.9\pi$.}
\label{fig:KI_finite_size}
\end{figure}

In deriving the asymptotic form of $\rho_{\rm s}(t)$ in equation~\eqref{eq:rho_S(t)_asymptotic} of the main text, we take the thermodynamic limit before the long-time limit. In a finite system, however, the reduced density matrix retains fluctuations around the infinite-temperature state even at late times, and the Bures distance fluctuates around a nonzero plateau. Figure~\ref{fig:KI_finite_size}\textbf{a} shows this behaviour for the kicked Ising chain with $L=10,12,\ldots,24$. For the larger systems, the Bures distance agrees with the exponential decay predicted by the dominant RP resonance after the initial transient and before reaching the plateau. Increasing $L$ lowers the plateau and extends the time interval over which the two curves agree.

The height of the plateau can be estimated by averaging over Haar-random pure states in the translation-invariant and reflection-symmetric sector, which contains the initial states and is preserved by the dynamics. The resulting mean distance, $D_{\rm Haar}$, is shown by the grey dashed lines in Fig.~\ref{fig:KI_finite_size}\textbf{a} and agrees well with the saturation values obtained from numerical simulation, particularly for the larger systems. Figure~\ref{fig:KI_finite_size}\textbf{b} shows that both $D_{\rm Haar}$ and the time-averaged saturation value $\overline D$ decrease exponentially with $L$.

This exponential scaling of the plateau height with $L$ also allows us to estimate how long the dominant RP resonance governs the decay of the Bures distance in finite systems. Writing as $D_{\rm RP}(t)\simeq A e^{-|\nu_{1,0}| t}$ the Bures distance predicted by the dominant RP resonant mode, we estimate the time $t_*$ at which this decay reaches the plateau by equating $D_{\rm RP}(t_*)$ with $\overline D(L)$, giving
\begin{align}
t_*\simeq\frac{\ln[A/\overline D(L)]}{|\nu_{1,0}|}
=O(L).
\label{eq:KI_time_window}
\end{align}
Figure~\ref{fig:KI_finite_size}\textbf{c} plots $t_*$ numerically evaluated from equation~\eqref{eq:KI_time_window} and shows that it increases approximately linearly with $L$. Thus, after contributions from faster-decaying modes have become negligible, $D(t)\simeq D_{\rm RP}(t)$ is expected to hold for $t\ll t_*=O(L)$. The proportionality coefficient between $t_*$ and $L$ depends on the details of the dynamics. 

\subsection{Cutoff convergence}

\begin{figure}[h]
\centering
\includegraphics[width=\textwidth]{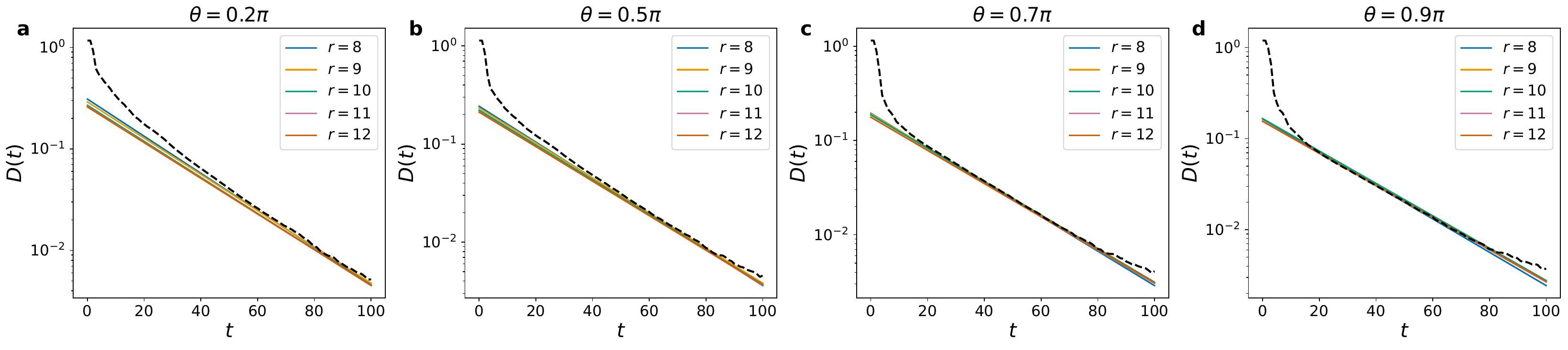}
\caption{\textbf{Convergence with the operator-support cutoff.} Coloured solid curves show the time evolution of the Bures distance for the kicked-Ising model calculated from the dominant RP resonance and its mode for $r=8,9,10,11,12$; black dashed curves show direct time evolution. The initial states are $\ket{\theta}=\cos(\theta/2)\ket{0}^{\otimes L}+\sin(\theta/2)\ket{1}^{\otimes L}$, with $\theta/\pi=0.2,0.5,0.7,0.9$ in panels \textbf{a}--\textbf{d}, respectively. We set $(J,\tau,h_x,h_z)=(1,0.65,0.9,0.8)$, $L=24$, and $\ell=4$.}
\label{fig:KI_cutoff}
\end{figure}

To check the dependence on support cutoff $r$, we recalculate the dominant RP resonance and its mode for $r=8,9,10,11,12$. Figure~\ref{fig:KI_cutoff} shows the time evolution of the Bures distance obtained by evaluating equation~\eqref{eq:D(t) large time KI TI} of the main text, with the dominant RP resonance and its mode calculated for several values of $r$. The curves become nearly indistinguishable as $r$ increases, indicating that the result is nearly converged at $r=12$ over the times shown. The late-time deviations between the predictions of the dominant RP resonance and the direct numerical results arise from the finite-size effects, as discussed in Sec.~\ref{sec:KI_finite_size}.

\subsection{Parameter dependence}
\label{sec:KI_parameter_cutoff}

\begin{figure}
\centering
\includegraphics[width=\textwidth]{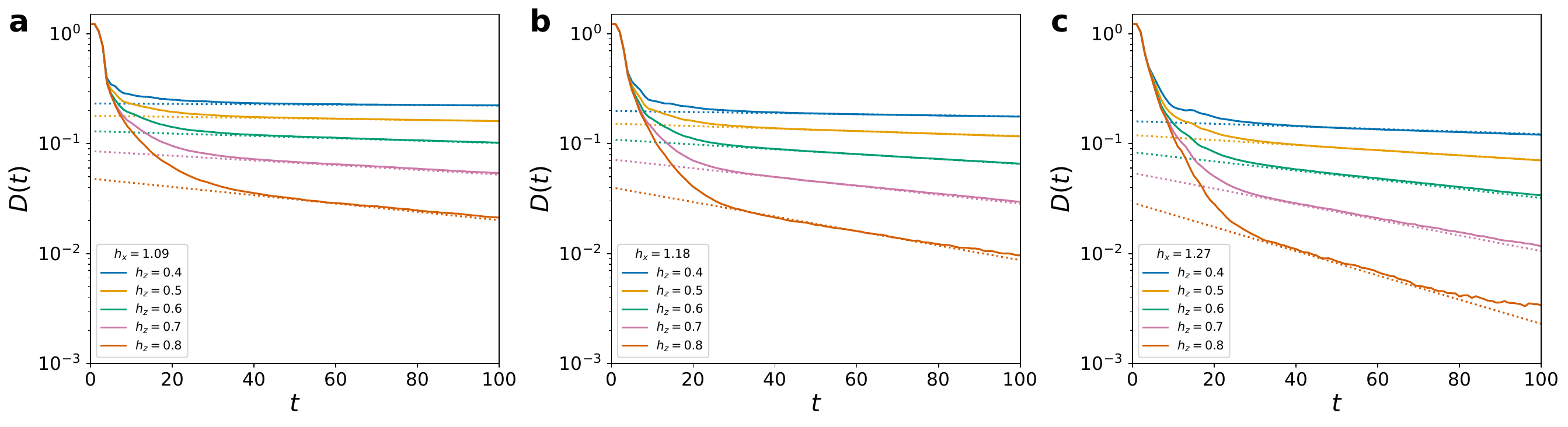}
\caption{\textbf{Local relaxation at other kicked Ising parameters.} Solid curves show the Bures distance from direct time evolution, and dotted curves show the distance calculated from the dominant RP resonance and its mode. We use $\tau=0.55$, $L=24$, $\ell=4$, $r=12$, periodic boundary conditions, and the initial product state $\ket{1}^{\otimes L}$.}
\label{fig:KI_parameters}
\end{figure}

Here, we test whether the dominant RP resonance also determines the decay of $D(t)$ at other parameter values. Figure~\ref{fig:KI_parameters} compares the Bures distance calculated from equation~\eqref{eq:D(t) large time KI TI} with that obtained from direct numerical simulations of the kicked Ising dynamics for several values of $(\tau,h_x,h_z)$ different from those used in Fig.~\ref{fig:D_c_fidelity} of the main text. It shows that the prediction of the RP resonances agrees well with the numerical result. The deviation for late times arises from the finite-size effects discussed in Sec.~\ref{sec:KI_finite_size}.

We also study $\tau=0.45$, at which the prethermalisation due to the quasi-conserved effective Hamiltonian, 
\begin{align}
H_{\rm eff}=H_x+H_z-\frac{\im\tau}{2}[H_x,H_z]-\frac{\tau^2}{12}\left([H_x,[H_x,H_z]]+[H_z,[H_z,H_x]]\right)+\cdots,
\label{eq:KI_Heff}
\end{align}
has been reported in Ref.~\cite{ref:Znidaric2024}. Here $H_x=-h_x\sum_j X_j$ and $H_z=-\sum_j(JZ_jZ_{j+1}+h_zZ_j)$, so that $U=e^{-\im\tau H_x}e^{-\im\tau H_z}$. The approximate conservation of $H_{\rm eff}$ suppresses heating towards infinite temperature and produces a long prethermal plateau. It is reflected in a dominant RP resonance $\lambda_{1,0}$ very close to, but smaller than, unity, accounting for the slow relaxation~\cite{ref:Znidaric2024}.

\begin{figure}
\centering
\includegraphics[width=0.8\textwidth]{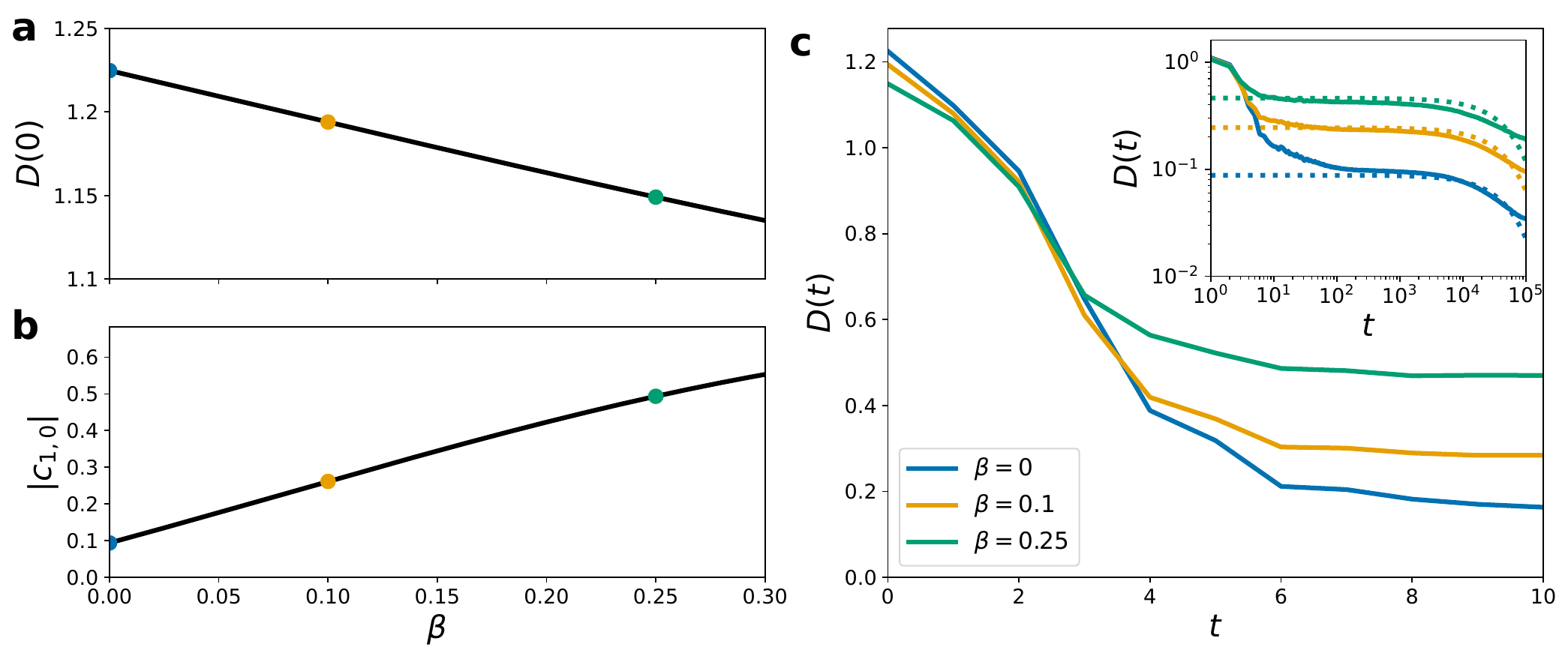}
\caption{\textbf{QME in the prethermal regime.} \textbf{a,} Initial Bures distance $D(0)$ for the states in equation~\eqref{eq:KI_prethermal_state}. \textbf{b,} Magnitude of the overlap $c_{1,0}$ with the dominant RP mode. \textbf{c,} Time evolution of the Bures distance for the states marked by coloured points in \textbf{a} and \textbf{b}. The inset shows the evolution over longer times; solid curves show direct time evolution, and dotted curves show equation~\eqref{eq:D(t) large time KI TI} of the main text, evaluated from the dominant RP resonance and its mode without fitting. We use $(\tau,h_x,h_z)=(0.45,0.9,0.8)$, $J=1$, $L=24$, $\ell=4$, $r=12$, and periodic boundary conditions.}
\label{fig:KI_prethermal}
\end{figure}
For the numerical benchmark, we consider the following initial state
\begin{align}
\ket{\psi_\beta}=e^{\im\beta\sum_j Z_jZ_{j+1}}
\left(\frac{\sqrt{3}}{2}\ket{0}-\frac{1}{2}\ket{1}\right)^{\otimes L}.
\label{eq:KI_prethermal_state}
\end{align}
Figure~\ref{fig:KI_prethermal}\textbf{a} shows that $D(0)$ decreases monotonically with $\beta$ over the range considered. Thus, for two initial states in equation~\eqref{eq:KI_prethermal_state} with $\beta_{\one}<\beta_{\two}$, state $\one$ is initially farther from equilibrium. For such a pair, inequality~\eqref{eq:QME_late}, together with equation~\eqref{eq:D(t) large time KI TI} of the main text, gives the necessary and sufficient condition for the QME, \mbox{$|c_{1,0}(\beta_{\one})|<|c_{1,0}(\beta_{\two})|$}. Since $|c_{1,0}|$ increases with $\beta$ for the selected states, as shown in Fig.~\ref{fig:KI_prethermal}\textbf{b}, this condition is satisfied and we expect the QME to occur. The time evolution of $D(t)$ in Fig.~\ref{fig:KI_prethermal}\textbf{c} indeed confirms this prediction. The inset further shows that the long-time decay agrees with equation~\eqref{eq:D(t) large time KI TI} of the main text, evaluated from the dominant RP resonance and its mode. The deviations at later times are due to finite-size effects, as also reported in Ref.~\cite{ref:Znidaric2024}.

\section{Ruelle-Pollicott resonances at the free-fermion point}
\label{sec:free_fermion}

In this section, we study subsystem equilibration and the QME in the kicked Ising model~\eqref{eq:KI} at $h_z=0$, where the dynamics reduces to that of a free fermion system.

\subsection{Time evolution of the correlation matrix}

Introducing the Majorana operators
\begin{align}
w_{j,0}=\left(\prod_{m<j}X_m\right)Z_j,
\qquad
w_{j,1}=-\left(\prod_{m<j}X_m\right)Y_j,
\qquad
\{w_{i,a},w_{j,b}\}=2\delta_{ij}\delta_{ab},
\label{eq:FF_majoranas}
\end{align}
the kicked Ising model at $h_z=0$ can be written in terms of quadratic Majorana operators as
\begin{align}
U=U_xU_z,
\qquad
U_x=e^{\tau h_x\sum_j w_{j,0}w_{j,1}},
\qquad
U_z=e^{\tau J\sum_j w_{j,1}w_{j+1,0}}.
\label{eq:FF_propagator}
\end{align}
In this case, the reduced density matrix is uniquely determined by the two-point correlation matrix $\Gamma(t)$, whose elements are given by 
\begin{align}
[\Gamma(t)]_{(i,a),(j,b)}
=\frac{\im}{2}\Tr\!\left[\rho(t)[w_{i,a},w_{j,b}]\right],
\qquad i,j\in{\rm S}.
\label{eq:FF_covariance}
\end{align}
We thus try to describe the time evolution of the two-point correlation matrix using the RP resonances. To this end, we introduce 
\begin{align}
B_d^{ab}(k)=\frac{\im}{2}\sum_j e^{-\im k(j+d/2)}[w_{j,a},w_{j+d,b}].
\label{eq:FF_bilinears}
\end{align}
In terms of $B_d^{ab}$, the two-point correlation matrix in the thermodynamic limit $L\to\infty$ can be written as 
\begin{align}
[\Gamma(t)]_{(i,a),(i+d,b)}
=\int_{-\pi}^{\pi}\frac{dk}{2\pi}e^{\im k(i+d/2)}
\Tr\!\left[\rho(0)\mathcal E^t[B_d^{ab}(k)]\right].
\label{eq:FF_covariance_from_bilinears}
\end{align}
For $d\geq0$, $B_d^{ab}(k)$ is the Fourier transform of a Pauli string supported on $d+1$ sites.

The quadratic form of equation~\eqref{eq:FF_propagator} implies that the Majorana operators transform linearly. 
Specifically, using equation~\eqref{eq:FF_bilinears}, one finds that 
\begin{align}\label{eq:Ew}
  \mathcal E[\boldsymbol w_j]=\sum_{l=-1}^1A_l\boldsymbol w_{j+l},
\end{align}
where $\boldsymbol w_j=(w_{j,0},w_{j,1})^T$ and
\begin{gather}
A_{-1}=\sin(2\tau J)
\begin{pmatrix}
0&-\cos(2\tau h_x)\\
0&\sin(2\tau h_x)
\end{pmatrix},\quad
A_0=\cos(2\tau J)
\begin{pmatrix}
\cos(2\tau h_x)&\sin(2\tau h_x)\\
-\sin(2\tau h_x)&\cos(2\tau h_x)
\end{pmatrix},\\
A_1=\sin(2\tau J)
\begin{pmatrix}
\sin(2\tau h_x)&0\\
\cos(2\tau h_x)&0
\end{pmatrix}.
\end{gather}
Substituting equation~\eqref{eq:Ew} into $\mathcal E_{k,r}[B_d^{ab}(k)]$, we obtain 
\begin{align}
\mathcal E_{k,r}[B_d^{ab}(k)]
=\sum_{l,m=-1}^1\sum_{c,e=0}^1
[A_l]_{ac}[A_m]_{be}e^{\im k(l+m)/2}B_{d-l+m}^{ce}(k).
\label{eq:FF_bilinear_evolution}
\end{align}
Here, the support cutoff $r$ restricts the operator basis to $\{B_d^{ab}\}_{d=0}^{r-1}$. The above equation shows that quadratic operators remain quadratic under time evolution.

We denote $B_d=(B_d^{00},B_d^{01},B_d^{10},B_d^{11})^T$ and write an eigenoperator of the truncated propagator as $\sum_d B_d^Tx_d$. The eigenvalue equation in this basis can be written as 
\begin{align}
\lambda x_d=\sum_{\delta=-2}^2 K_\delta(k)^T x_{d-\delta},
\qquad
K_\delta(k)=\sum_{m-l=\delta}e^{\im k(l+m)/2}A_l\otimes A_m.
\label{eq:FF_recurrence}
\end{align}
The support cutoff requires $x_d=0$ for $d\ge r$. At the other boundary, the identities $B_{-d}^{ab}(k)=-B_d^{ba}(k)$ imply $x_{-d}^{ab}=-x_d^{ba}$, where $x_d^{ab}$ is the coefficient of $B_d^{ab}(k)$. In particular, $x_0^{00}=x_0^{11}=0$ and $x_0^{10}=-x_0^{01}$. In the limit $r\to\infty$, the effects of these boundary conditions can be neglected in the bulk, $1\ll d\ll r$. Substituting the plane-wave form $x_d=e^{\im qd}v$ into equation~\eqref{eq:FF_recurrence} then gives the eigenvalue equation in Fourier space,
\begin{align}
\lambda v
=\left[\sum_{\delta=-2}^2e^{-\im q\delta}K_\delta(k)^T\right]v
=\left[R(k/2+q)^T\otimes R(k/2-q)^T\right]v,
\label{eq:FF_fourier_eigenvalue}
\end{align}
where
\begin{align}
R(q):=\sum_{l=-1}^1e^{\im ql}A_l
=
\begin{pmatrix}
\cos(2\tau h_x)&\sin(2\tau h_x)\\
-\sin(2\tau h_x)&\cos(2\tau h_x)
\end{pmatrix}
\begin{pmatrix}
\cos(2\tau J)&-e^{-\im q}\sin(2\tau J)\\
e^{\im q}\sin(2\tau J)&\cos(2\tau J)
\end{pmatrix}.
\label{eq:FF_R}
\end{align}
The eigenvalues of $R(q)$ are $e^{\pm\im\varepsilon(q)}$, with
\begin{align}
\varepsilon(q)=\arccos\!\left[\cos(2\tau h_x)\cos(2\tau J)
+\sin(2\tau h_x)\sin(2\tau J)\cos q\right].
\label{eq:FF_dispersion}
\end{align}
Hence, we obtain the eigenvalues of the truncated propagator in the limit $r\to\infty$ as 
\begin{align}
\lambda_{\mu\nu,k}(q)
=e^{\im\mu\varepsilon(q+k/2)-\im\nu\varepsilon(q-k/2)},
\qquad \mu,\nu=\pm1.
\label{eq:FF_spectrum}
\end{align}
Equation~\eqref{eq:FF_spectrum} shows that the eigenvalues of the truncated propagator arrive at the unit circle, $|\lambda_{\mu\nu,k}(q)|=1$, in the limit $r\to\infty$. This result is consistent with the fact that there exist infinitely many conserved quantities due to the integrability at the free-fermion point.

Substituting these modes into equation~\eqref{eq:FF_covariance_from_bilinears} gives the time evolution of the correlation matrix. For a translationally invariant initial state, only $k=0$ contributes, and we find
\begin{align}
\Gamma(t)_{i,j}
=\int_{-\pi}^{\pi}\frac{dq}{2\pi}e^{\im q(i-j)}\mathcal G(q,t),
\quad 
\mathcal G(q,t)
=\sum_{\mu,\nu=\pm1}\lambda_{\mu\nu,0}(q)^t
\Xi_\mu(q)\mathcal G_0(q)\Xi_\nu(q).
\label{eq:FF_symbol_spectral}
\end{align}
Here $\Xi_\mu(q)$ is the projector onto the eigenspace of $R(q)$ with eigenvalue $e^{\im\mu\varepsilon(q)}$ and $\mathcal G_0(q)=\sum_d e^{\im qd}\Gamma_{i,i+d}(0)$. Equation~\eqref{eq:FF_symbol_spectral} reproduces the exact Majorana two-point correlation matrix obtained by directly diagonalizing the Floquet operator of the kicked Ising chain \cite{ref:Collura2022}, implying that the exact subsystem dynamics for Gaussian initial states is recovered from the framework based on RP resonances.

\subsection{QME in the kicked Ising model at the free-fermion point}

We now investigate the QME at the free-fermion point using the two-point correlation matrix in equation~\eqref{eq:FF_symbol_spectral}. We choose as initial states the ground states of the XY Hamiltonian (whose basis is rotated)
\begin{align}
H_0(\gamma,h)
=-\sum_j\left[\frac{1+\gamma}{2}Z_jZ_{j+1}
+\frac{1-\gamma}{2}Y_jY_{j+1}+hX_j\right].
\label{eq:FF_initial_Hamiltonian}
\end{align}
The ground state can be readily obtained by diagonalizing the above Hamiltonian using the Jordan-Wigner and the Bogoliubov transformations~(see e.g. \cite{fagotti2008}). For this initial state, $\mathcal{G}_0(q)$ in equation~\eqref{eq:FF_symbol_spectral} reads 
\begin{align}
\mathcal G_0(q)=
\begin{pmatrix}0&f(q)\\-f(q)^*&0\end{pmatrix},
\qquad
f(q)=\frac{\cos q-h-\im\gamma\sin q}
{\sqrt{(\cos q-h)^2+\gamma^2\sin^2q}}.
\label{eq:FF_initial_symbol}
\end{align}

Following Ref.~\cite{ref:Ares2025}, we quantify relaxation by the relative entropy
\begin{align}
S\bigl(\rho_{\rm s}(t)\Vert\rho_{\rm s}(\infty)\bigr)
=\Tr\!\left[\rho_{\rm s}(t)\left\{\log\rho_{\rm s}(t)-\log\rho_{\rm s}(\infty)\right\}\right].
\label{eq:FF_relative_entropy}
\end{align}
Each initial state is compared with its own stationary state, obtained by retaining the $\mu=\nu$ terms in equation~\eqref{eq:FF_symbol_spectral}, with symbol $\mathcal G_\infty(q)=\sum_{\mu=\pm1}\Xi_\mu(q)\mathcal G_0(q)\Xi_\mu(q)$. The relative entropy can be written in terms of the correlation matrices as~\cite{ref:FinsterMuch2025}
\begin{align}
S\bigl(\rho_{\rm s}(t)\Vert\rho_{\rm s}(\infty)\bigr)
=\Tr\left[
\frac{I+\im\Gamma(t)}{2}
\left\{
\log\!\left(\frac{I+\im\Gamma(t)}{2}\right)
-\log\!\left(\frac{I+\im\Gamma(\infty)}{2}\right)
\right\}\right].
\label{eq:FF_relative_entropy_covariance}
\end{align}

\begin{figure}[htb]
\centering
\includegraphics[width=0.5\textwidth]{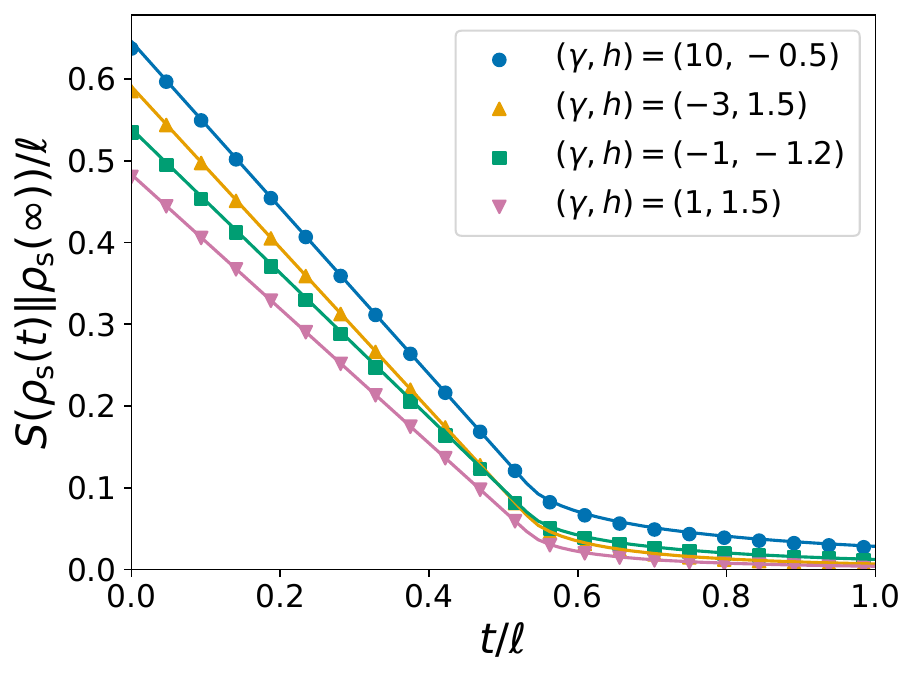}
\caption{\textbf{QME at the free-fermion point.} The symbols show the exact values of relative entropy calculated using equation~\eqref{eq:FF_relative_entropy_covariance} for initial ground states of equation~\eqref{eq:FF_initial_Hamiltonian}. The solid curves show the analytical expression~\eqref{eq:FF_entropy_quasiparticle}. We use $(\tau,h_x,h_z)=(0.6,1.1,0)$, $J=1$, and $\ell=128$ in the thermodynamic limit.}
\label{fig:FF_QME}
\end{figure}

When the dynamics is local, the relative entropy in equation~\eqref{eq:FF_relative_entropy} for a large subsystem can be approximated by the entropy difference~\cite{ref:Ares2025},
\begin{align}
S(\rho_{\rm s}(t)\Vert\rho_{\rm s}(\infty))
\simeq S[\rho_{\rm s}(\infty)]-S[\rho_{\rm s}(t)],
\label{eq:FF_entropy_deficit}
\end{align}
where $S[\rho]=-\Tr[\rho\log \rho]$ is the von Neumann entropy. 
The entropy difference in equation~\eqref{eq:FF_entropy_deficit} can be calculated analytically using the quasiparticle picture, a semiclassical theory describing the dynamics of integrable systems~\cite{ref:Ares2025}. In this picture, switching on the Floquet time evolution at $t=0$ excites pairs of entangled quasiparticles, which propagate ballistically in opposite directions. The entropy difference is determined by the number of pairs whose two members remain inside the subsystem, with each mode weighted by the Yang-Yang entropy. Counting these pairs in the scaling limit $\ell,t\to\infty$ at fixed $t/\ell$, we obtain
\begin{align}
S(\rho_{\rm s}(t)\Vert\rho_{\rm s}(\infty))\simeq
\int_{-\pi}^{\pi}\frac{dq}{2\pi}
\max\!\left(\ell-2t|v(q)|,0\right)s[n(q)],
\label{eq:FF_entropy_quasiparticle}
\end{align}
where $s(x)=-x\log x-(1-x)\log (1-x)$, $v(q)=\partial_q\varepsilon(q)$ is the group velocity and $n(q)$ is the occupation of the Floquet quasiparticle mode, given by
\begin{align}
n(q)=\frac{1+\im\,\mathrm{tr}[\Xi_+(q)\mathcal G_0(q)]}{2}.
\label{eq:FF_occupation}
\end{align}
Figure~\ref{fig:FF_QME} compares the relative entropy evaluated using equation~\eqref{eq:FF_relative_entropy_covariance} with the analytical expression~\eqref{eq:FF_entropy_quasiparticle} for several initial states. The two calculations agree well and exhibit the QME for some pairs of the initial configurations. 

The QME at the free-fermion point can be understood within the quasiparticle picture. According to equation~\eqref{eq:FF_entropy_quasiparticle}, quasiparticles of all modes contribute to the entropy difference at $t=0$. At late times, however, the quasiparticle pairs propagate and escape from the subsystem, and only the slowest modes with $|v(q)|<\ell/(2t)$ contribute to the entropy difference. The factor $\max(\ell-2|v(q)|t,0)$ in equation~\eqref{eq:FF_entropy_quasiparticle} vanishes for fast-propagating modes with $|v(q)|>\ell/(2t)$ and therefore excludes their contributions. Thus, an initial state with a larger total entropy difference but a smaller contribution from the slowest modes can start farther from its stationary state yet relax faster, giving rise to the QME~\cite{ref:Rylands2024}.

Equation~\eqref{eq:FF_occupation} shows that the occupations $n(q)$ are determined by $\mathrm{tr}[\Xi_+(q)\mathcal G_0(q)]$, the initial state's overlap with the $\mu=\nu=+1$ mode in equation~\eqref{eq:FF_symbol_spectral}. The entropy weights $s[n(q)]$ that control the contribution of the slowest quasiparticles are therefore fixed by the initial state's overlaps with RP resonant modes. Thus, the established quasiparticle picture for the QME in integrable systems is consistent with our conclusion that the initial state's overlaps with the dominant RP resonant modes determine the QME.

\section{Ruelle-Pollicott resonances at the dual-unitary point}\label{sec:DU}
Here, we study the kicked Ising model at the dual-unitary point, $J=h_x=1$ and $\tau=\pi/4$~\cite{ref:Bertini2026}. At this point, two periods of the Floquet evolution of the kicked Ising model can be rewritten as a brickwork circuit constructed from the two-site gate
\begin{align}
v=e^{\im\pi Z\otimes Z/4}
\left[\left(e^{\im\phi Z/2}e^{\im\pi X/4}e^{\im\phi Z/2}\right)
\otimes e^{\im\pi X/4}\right]e^{\im\pi Z\otimes Z/4},
\qquad \phi=2\tau h_z.
\label{eq:DU_gate}
\end{align}
The gate $v$ is unitary both in the time direction and after interchanging space and time. Defining one full period of the brickwork circuit by
\begin{align}
V=\left(\prod_{j\,\mathrm{odd}}v_{j,j+1}\right)
  \left(\prod_{j\,\mathrm{even}}v_{j,j+1}\right),
\qquad
U^2=U_{\rm o}^{\dagger}VU_{\rm o},
\qquad
U_{\rm o}=\prod_{j\,\mathrm{odd}}
e^{\im\pi Z_jZ_{j+1}/4}e^{\im\phi Z_j/2},
\label{eq:DU_frame}
\end{align}
we analyse the dynamics generated by $V$ in the thermodynamic limit. In this section, $t$ counts full periods of $V$, each consisting of two layers and corresponding to two periods of $U$. 

\subsection{Operator spreading and the truncated propagator}
The circuit generated by $V$ in equation~\eqref{eq:DU_frame} is invariant under two-site translations, i.e., $s=2$. To describe the time evolution of operators, we first distinguish Pauli strings according to the parities of their two endpoints. The support length of a nonidentity Pauli string is the length of the smallest interval containing all its nonidentity factors, including any identity factors between them. If this length is odd, the endpoints have the same parity; if it is even, they have opposite parities.

Let $\mathscr P^{\rm e}$ and $\mathscr P^{\rm o}$ denote the sets of Pauli strings whose leftmost nonidentity factor is at site $0$ and whose support lengths are even and odd, respectively. We define four operator subspaces,
\begin{equation}
\begin{aligned}
\mathcal H_{\rm ee}&=\mathrm{span}\{\mathcal T^{2n}P\mid P\in\mathscr P^{\rm e},\ n\in\mathbb Z\},\\
\mathcal H_{\rm eo}&=\mathrm{span}\{\mathcal T^{2n}P\mid P\in\mathscr P^{\rm o},\ n\in\mathbb Z\},\\
\mathcal H_{\rm oo}&=\mathrm{span}\{\mathcal T^{2n+1}P\mid P\in\mathscr P^{\rm o},\ n\in\mathbb Z\},\\
\mathcal H_{\rm oe}&=\mathrm{span}\{\mathcal T^{2n+1}P\mid P\in\mathscr P^{\rm e},\ n\in\mathbb Z\}.
\end{aligned}
\label{eq:DU_operator_subspaces}
\end{equation}
The first index in $\mathcal{H}_{ab}$ specifies the parity of the leftmost site, and the second specifies the parity of the support length. Two-site translations preserve both indices, so each subspace can be resolved into momentum sectors independently.

\begin{figure}[htb]
\centering
\includegraphics[width=0.4\textwidth]{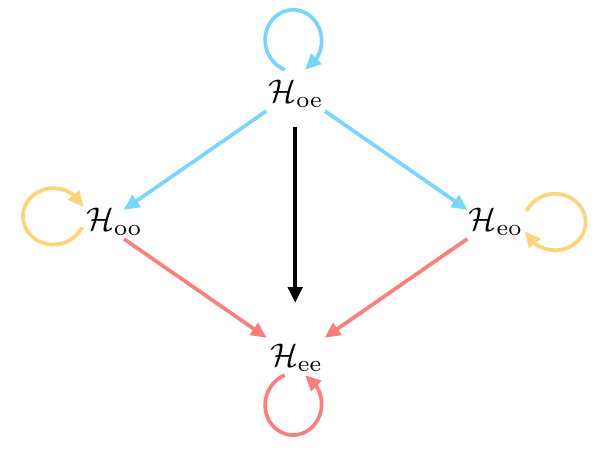}
\caption{\textbf{Operator spreading at the dual-unitary point.} Arrows show the allowed transitions between the subspaces in equation~\eqref{eq:DU_operator_subspaces} under $\mathcal E(P)=V^\dagger PV$. Blue, red, and yellow arrows indicate a decrease, an increase, and no change in support length, respectively. For the transition indicated by the black arrow, the support length may increase, decrease, or remain unchanged.}
\label{fig:DU_operator_flow}
\end{figure}

Dual unitarity constrains the motion of the two ends of a Pauli string under $\mathcal E(P)=V^\dagger PV$~\cite{ref:Bertini2026}. Figure~\ref{fig:DU_operator_flow} summarises the resulting transitions between the four subspaces. The red loop at $\mathcal H_{\rm ee}$ shows that, for $P\in\mathcal H_{\rm ee}$, every output Pauli string remains in this subspace and has a strictly larger support length. For $P\in\mathcal H_{\rm eo}$ or $\mathcal H_{\rm oo}$, an output Pauli string either remains in the same subspace with the same support length (yellow loops) or enters $\mathcal H_{\rm ee}$ with a larger support (red arrows). There is no transition between $\mathcal H_{\rm eo}$ and $\mathcal H_{\rm oo}$.

The blue arrows show that, for $P\in\mathcal H_{\rm oe}$, an output Pauli string can remain in $\mathcal H_{\rm oe}$ or enter $\mathcal H_{\rm eo}$ or $\mathcal H_{\rm oo}$, in each case with a strictly smaller support length. It can also enter $\mathcal H_{\rm ee}$ along the black arrow, in which case its support length may increase, decrease, or remain unchanged. Once an operator component enters $\mathcal H_{\rm ee}$, its support subsequently grows and it cannot return to the other subspaces.

The directions of the arrows in Fig.~\ref{fig:DU_operator_flow} determine which blocks of the truncated propagator can be nonzero. At fixed momentum $k$, we retain only Pauli strings with support length at most $r$ and order the basis by the subspaces $(\mathcal H_{\rm ee},\mathcal H_{\rm eo},\mathcal H_{\rm oo},\mathcal H_{\rm oe})$. Since the transitions between different subspaces proceed only towards earlier entries in this ordering, the truncated propagator takes the block upper triangular form
\begin{align}
\mathcal E_{k,r}=
\begin{pmatrix}
M_{\rm ee}&M_{{\rm eo}\to{\rm ee}}&M_{{\rm oo}\to{\rm ee}}&M_{{\rm oe}\to{\rm ee}}\\
0&M_{\rm eo}&0&M_{{\rm oe}\to{\rm eo}}\\
0&0&M_{\rm oo}&M_{{\rm oe}\to{\rm oo}}\\
0&0&0&M_{\rm oe}
\end{pmatrix}.
\label{eq:DU_block}
\end{align}
The block $M_{ab}$ describes propagation within $\mathcal H_{ab}$, while $M_{ab\to cd}$ describes propagation from $\mathcal H_{ab}$ to $\mathcal H_{cd}$. Within each subspace, ordering the Pauli strings by increasing support length makes $M_{\rm ee}$ strictly lower triangular and $M_{\rm oe}$ strictly upper triangular. Repeated application of $M_{\rm ee}$ eventually increases the support beyond the cutoff, whereas repeated application of $M_{\rm oe}$ eventually leaves no admissible Pauli string in that subspace. Thus, for some finite integer $N$, $M_{\rm ee}^N=M_{\rm oe}^N=0$, implying that these two blocks are nilpotent and have only zero eigenvalues.

Since the eigenvalues of a block triangular matrix are those of its diagonal blocks, every nonzero eigenvalue of $\mathcal E_{k,r}$ comes from $M_{\rm eo}$ or $M_{\rm oo}$. The yellow loops in Fig.~\ref{fig:DU_operator_flow} indicate that propagation within either of these subspaces preserves the support length, so these blocks further decompose as
\begin{align}
M_a=\bigoplus_{\substack{1\leq\zeta\leq r\\\zeta:\mathrm{odd}}}M_{a,\zeta},
\qquad a=\mathrm{eo},\mathrm{oo}.
\label{eq:DU_support_blocks}
\end{align}
The block $M_{a,\zeta}$ acts on Pauli strings of support length exactly $\zeta$ in $\mathcal H_a$. Its matrix elements are independent of $r$ once $r\geq\zeta$, because the propagation represented by this block does not change the support length. Its eigenvalues strictly inside the unit circle therefore give exact RP resonances without extrapolating in $r$.

The off-diagonal blocks in equation~\eqref{eq:DU_block} determine the remaining components of the corresponding eigenoperators. To calculate them explicitly, let $\lambda\ne0$ be an eigenvalue of $M_{{\rm eo},\zeta}$, with right and left eigenvectors $\boldsymbol u_\zeta^{\rm R}$ and $\boldsymbol u_\zeta^{\rm L}$. Solving $\mathcal E_{k,r}\boldsymbol v^{\rm R}=\lambda\boldsymbol v^{\rm R}$ and $\mathcal E_{k,r}^\dagger\boldsymbol v^{\rm L}=\lambda^*\boldsymbol v^{\rm L}$ gives
\begin{equation}
\boldsymbol v^{\rm R}=
\begin{pmatrix}
(\lambda I-M_{\rm ee})^{-1}M_{{\rm eo}\to{\rm ee}}\boldsymbol u_\zeta^{\rm R}\\
\boldsymbol u_\zeta^{\rm R}\\
0\\
0
\end{pmatrix},
\qquad
\boldsymbol v^{\rm L}=
\begin{pmatrix}
0\\
\boldsymbol u_\zeta^{\rm L}\\
0\\
(\lambda^* I-M_{\rm oe}^\dagger)^{-1}M_{{\rm oe}\to{\rm eo}}^\dagger\boldsymbol u_\zeta^{\rm L}
\end{pmatrix}.
\label{eq:DU_full_eigenvectors}
\end{equation}
Using $M_{\rm ee}^N=M_{\rm oe}^N=0$, we expand the first component of $\boldsymbol v^{\rm R}$ and the fourth of $\boldsymbol v^{\rm L}$ in equation~\eqref{eq:DU_full_eigenvectors} as
\begin{align}
(\lambda I-M_{\rm ee})^{-1}M_{{\rm eo}\to{\rm ee}}\boldsymbol u_\zeta^{\rm R}
&=\sum_{n=0}^{N-1}\frac{M_{\rm ee}^n}{\lambda^{n+1}}M_{{\rm eo}\to{\rm ee}}\boldsymbol u_\zeta^{\rm R},
\label{eq:DU_right_eigenvector_tail}\\
(\lambda^* I-M_{\rm oe}^\dagger)^{-1}M_{{\rm oe}\to{\rm eo}}^\dagger\boldsymbol u_\zeta^{\rm L}
&=\sum_{n=0}^{N-1}\frac{(M_{\rm oe}^\dagger)^n}{(\lambda^*)^{n+1}}M_{{\rm oe}\to{\rm eo}}^\dagger\boldsymbol u_\zeta^{\rm L}.
\label{eq:DU_left_eigenvector_tail}
\end{align}
The coupling $M_{{\rm eo}\to{\rm ee}}$ takes a Pauli string of length $\zeta$ to Pauli strings of strictly larger support, and every subsequent application of $M_{\rm ee}$ increases their support further. Hence equation~\eqref{eq:DU_right_eigenvector_tail} means that the right eigenoperator corresponding to $\boldsymbol u_\zeta^{\rm R}$ consists of its length-$\zeta$ component in $\mathcal H_{\rm eo}$ and longer Pauli strings in $\mathcal H_{\rm ee}$. For the left eigenoperator corresponding to $\boldsymbol u_\zeta^{\rm L}$, $M_{{\rm oe}\to{\rm eo}}^\dagger$ and $M_{\rm oe}^\dagger$ in equation~\eqref{eq:DU_left_eigenvector_tail} increase the support length, as they respectively reverse propagation from $\mathcal H_{\rm oe}$ to $\mathcal H_{\rm eo}$ and from $\mathcal H_{\rm oe}$ to $\mathcal H_{\rm oe}$, both of which decrease support length (see Fig.~\ref{fig:DU_operator_flow}). Therefore, the same argument as for the right eigenoperator applies, and the left eigenoperator corresponding to $\boldsymbol u_\zeta^{\rm L}$ contains no Pauli strings with support lengths smaller than $\zeta$.

This restriction determines which RP resonances can contribute to the subsystem equilibration. For $s=2$, the reduced mode in equation~\eqref{eq:main_integral} of the main text is
\begin{align}
\varrho_{\alpha,k}
=\frac{1}{2^\ell}\sum_{\substack{n\in\mathbb Z,\ P\in\mathscr P_{\ell,2}\\\operatorname{supp}(\mathcal T^{2n}P)\subseteq\mathrm S}}
e^{\im kn}(v_{\alpha,P_k}^{\rm L})^*\mathcal T^{2n}P.
\label{eq:DU_reduced_mode_general}
\end{align}
Note that only the Pauli strings that fit entirely inside the subsystem $\mathrm S$ can appear in the above equation. If a mode originates from $M_{a,\zeta}$ with $\zeta>\ell$, all Pauli strings in its left eigenoperator are too long to fit, and $\varrho_{\alpha,k}=0$. Such a mode cannot contribute to the reduced density matrix. The relevant nonzero RP resonances are therefore contained in the blocks with $\zeta\leq\ell$, and all these blocks are already present at $r=\ell$. Moreover, the finite sums in equations~\eqref{eq:DU_right_eigenvector_tail} and~\eqref{eq:DU_left_eigenvector_tail} only generate longer Pauli strings, so increasing the cutoff beyond $\ell$ cannot change the left-eigenoperator components that enter $\varrho_{\alpha,k}$. Hence, setting $r=\ell$ is sufficient to obtain exactly all nonzero RP resonances that contribute to a subsystem of length $\ell$. The initial overlaps $\Tr_{}[\rho(0)V_{\alpha,k}^{\rm R}]$ can still involve longer Pauli strings in the right eigenoperators, including those with support lengths greater than $\ell$.

\subsection{Exact one-site reduced dynamics}
At the dual-unitary point, these constraints on operator spreading allow us to calculate the time evolution of a one-site reduced density matrix analytically for suitable initial states. For $\ell=1$, only the truncated propagator in equation~\eqref{eq:DU_support_blocks} with $\zeta=1$ contributes to the RP resonances relevant to the subsystem equilibration. Direct conjugation by the gate in equation~\eqref{eq:DU_gate} gives, for an odd site,
\begin{align}
\begin{aligned}
V^{\dagger}X_1V&=\cos^2\phi\,X_3+O_X,\\
V^{\dagger}Y_1V&=-\cos\phi\sin\phi\,X_3+O_Y,\\
V^{\dagger}Z_1V&=O_Z,
\end{aligned}
\label{eq:DU_local_evolution}
\end{align}
where $O_X,O_Y,O_Z\in\mathcal H_{\rm ee}$ are linear combinations of Pauli strings acting nontrivially on more than one site. From equation~\eqref{eq:DU_local_evolution}, we obtain 
\begin{align}
M_{{\rm eo},1}(k)=e^{-\im k}
\begin{pmatrix}
\cos^2\phi&0&0\\
\cos\phi\sin\phi&0&0\\
0&0&0
\end{pmatrix},
\qquad
M_{{\rm oo},1}(k)=e^{\im k}
\begin{pmatrix}
\cos^2\phi&-\cos\phi\sin\phi&0\\
0&0&0\\
0&0&0
\end{pmatrix}.
\label{eq:DU_one_site_blocks}
\end{align}
The phases $e^{\mp\im k}$ describe translation by one unit cell in opposite directions.

By diagonalizing the matrices in equation~\eqref{eq:DU_one_site_blocks}, we obtain the two nonzero RP resonances as
\begin{align}
\lambda_{{\rm eo},k}=e^{-\im k}\cos^2\phi,
\qquad
\lambda_{{\rm oo},k}=e^{\im k}\cos^2\phi.
\label{eq:DU_resonances}
\end{align}
They remain eigenvalues of the truncated propagator for every $r\geq1$. For an odd-site subsystem $\mathrm S=\{1\}$, the right and left eigenvectors of $M_{{\rm oo},1}(k)$ can be chosen as $(1,0,0)^T$ and $(1,-\tan\phi,0)^T$, respectively. The one-site component of the right eigenoperator is therefore $X_k=\sum_j e^{-\im kj}X_{2j+1}$, while $\varrho_{\mathrm{oo},k}$ reads 
\begin{align}
\varrho_{{\rm oo},k}=\frac{X-Y\tan\phi}{2}.
\label{eq:DU_reduced_mode}
\end{align}
Thus, the nonzero decay modes visible to one site all have modulus $\cos^2\phi$, and their weights are determined by the initial overlaps with the corresponding right eigenoperators.

To evaluate the initial overlaps explicitly, we choose the initial state
\begin{align}
\rho(0)=\bigotimes_j\left(\frac{I_{2j}}{2}\otimes\sigma_{2j+1}\right),
\qquad
\sigma_{2j+1}=\sigma=\frac{I+xX+yY+zZ}{2}.
\label{eq:DU_initial}
\end{align}
All even sites are maximally mixed, and all odd sites are prepared in the same state. With the normalization chosen above, the support-one component of $V_{{\rm oo},k}^{\rm R}$ is $X_k$. Its remaining components are generated by the blocks $M_{{\rm oo}\to{\rm ee}}$ and $M_{\rm ee}$ in equation~\eqref{eq:DU_block}, so $V_{{\rm oo},k}^{\rm R}-X_k$ consists entirely of Fourier-transformed Pauli strings in $\mathcal H_{\rm ee}$. Each such Pauli string has a nonidentity Pauli operator at its even left endpoint. The factor at the even left endpoint vanishes because a nonidentity Pauli operator has zero expectation value in the maximally mixed state. We thus obtain
\begin{align}
\Tr[\rho(0)V_{{\rm oo},k}^{\rm R}]
=\Tr[\rho(0)X_k]
=\sum_{j\in\mathbb Z}e^{-\im kj}\Tr(\sigma X)
=x\sum_{j\in\mathbb Z}e^{-\im kj}
=2\pi x\,\delta(k).
\label{eq:DU_overlap}
\end{align}
Substituting equations~\eqref{eq:DU_resonances}, \eqref{eq:DU_reduced_mode}, and \eqref{eq:DU_overlap} into equation~\eqref{eq:main_integral} of the main text, we arrive at 
\begin{align}
\rho_{\rm s}(t)=\frac{I}{2}
+\frac{x}{2}\left[\cos^{2t}\!\phi\,X
-\cos^{2t-1}\!\phi\sin\phi\,Y\right],
\qquad t\geq1,
\label{eq:DU_rho_exact}
\end{align}
with $\rho_{\rm s}(0)=\sigma$. We note that equation~\eqref{eq:DU_rho_exact} can be expressed as the repeated application of the same one-site quantum channel,
\begin{align}
\Phi_\phi(\sigma)
=\frac{I}{2}+\frac{\Tr(\sigma X)}{2}
\left(\cos^2\phi\,X-\cos\phi\sin\phi\,Y\right),
\qquad
\rho_{\rm s}(t)=\Phi_\phi^t(\sigma).
\label{eq:DU_channel}
\end{align}
This channel is completely positive and trace preserving, implying that the reduced density matrix follows Markovian dynamics. 

The Markovian evolution in equation~\eqref{eq:DU_channel} follows from two properties. First, dual unitarity prevents the terms that have spread beyond one site in equation~\eqref{eq:DU_local_evolution} from returning to one-site operators. Thus, the one-site operator components evolve without feedback from the larger-support terms generated at earlier times. Second, for the initial state in equation~\eqref{eq:DU_initial}, $\Tr_{}[\rho(0)V_{{\rm oo},k}^{\rm R}]$ in equation~\eqref{eq:DU_overlap} reduces to $2\pi\delta(k)\Tr[\rho_{\rm s}(0)X]$, so the weight of the relaxation mode is determined entirely by the initial reduced density matrix. 

In the ordered operator basis $\{I,X,Y,Z\}$, the matrix representation of $\Phi_\phi$ is
\begin{equation}\label{eq:DU_channel_matrix}
[\Phi_\phi]=
\begin{pmatrix}
1&0&0&0\\
0&\cos^2\phi&0&0\\
0&-\cos\phi\sin\phi&0&0\\
0&0&0&0
\end{pmatrix}.
\end{equation}
Its eigenvalues are $1,\cos^2\phi,0,0$. Note that the eigenvalue $\cos^2\phi$, which determines the relaxation rate in equation~\eqref{eq:DU_channel}, coincides with the RP resonances in equation~\eqref{eq:DU_resonances} at $k=0$, selected by the two-site translation symmetry of the initial state in equation~\eqref{eq:DU_initial}. In addition, its left and right eigenoperators
\begin{equation}\label{eq:DU_channel_eigenoperators}
L_\phi=X,\qquad R_\phi=X-Y\tan\phi,
\end{equation}
also coincide with the one-site components of $V_{{\rm oo},0}^{\rm R}$ and $V_{{\rm oo},0}^{\rm L}$, respectively. Note that the operator and density-matrix propagators are adjoints, and hence the right eigenoperator of $\Phi_\phi$ corresponds to the left eigenoperator of the RP resonance, and vice versa. Thus, in the present setting, the RP resonance and its mode directly reduce to the Liouvillian spectrum governing Markovian dynamics.

These results also yield an explicit condition for the QME. Using equation~\eqref{eq:DU_channel}, we find that the squared Hilbert--Schmidt distance from the equilibrium state $I/2$ is given by $D_{\rm HS}^2(0)=(x^2+y^2+z^2)/2$ and $D_{\rm HS}^2(t)=x^2|\cos\phi|^{4t-2}/2$ for $t\geq1$. Thus, the QME occurs if the initial state with the larger $x^2+y^2+z^2$ has the smaller $|x|$, which is the weight of the RP resonant mode $V_{{\rm oo},k}^{\rm R}$ at $k=0$ in the global initial state or, equivalently, the overlap of the initial state of the subsystem with the Liouvillian eigenoperator $L_\phi$.

\section{Quantum automaton circuit}\label{sec:QCA}
We consider the QAC shown in Fig.~\ref{fig:strong_QME}\textbf{b} of the main text. One Floquet period consists of two staggered layers of the same four-site gate, composed of CNOT and Toffoli gates, with the two layers shifted by two sites. The dynamics is invariant under translation by $s=4$ sites. Since each gate permutes computational-basis states without introducing phases, the time evolution takes the form
\begin{align}
U^t\ket{x}=\ket{f_t(x)},
\label{eq:QCA_map}
\end{align}
where $x$ is a bit configuration and $f_t$ is a bijection. In this section, we show that this property allows us both to simulate local equilibration directly in the thermodynamic limit and to simplify the calculation of RP resonances for the QAC.

\subsection{Hilbert--Schmidt distance and Monte Carlo simulation}
For this circuit, we quantify equilibration by the Hilbert--Schmidt distance from $\rho_{\rm s}^{\rm eq}=I/2^\ell$. Its square is given by
\begin{align}
D_{\rm HS}^2(t)=\Tr\!\left[\left(\rho_{\rm s}(t)-\frac{I}{2^\ell}\right)^2\right]
=\Tr[\rho_{\rm s}(t)^2]-2^{-\ell}.
\label{eq:QCA_HS}
\end{align}
Thus, it suffices to calculate the subsystem purity. This can be done by Monte Carlo sampling. To describe the method, we first consider a finite chain and write $p(a,a')=\bra{a}\rho(0)\ket{a'}$, where $\ket{a}$ and $\ket{a'}$ are computational-basis states. The purity is the expectation value of the operator $\mathrm{SWAP}_{\rm s}$ that exchanges the subsystem configurations between two replicas:
\begin{align}
\Tr[\rho_{\rm s}(t)^2]
=\Tr[\rho(t)^{\otimes2}\mathrm{SWAP}_{\rm s}]
=\sum_{x,y}p\!\left(f_t^{-1}(x),f_t^{-1}(\widetilde{x})\right)
 p\!\left(f_t^{-1}(y),f_t^{-1}(\widetilde{y})\right),
\label{eq:QCA_purity}
\end{align}
where $x=x_{\rm s}x_{\rm e}$ and $y=y_{\rm s}y_{\rm e}$, while $\widetilde{x}=y_{\rm s}x_{\rm e}$ and $\widetilde{y}=x_{\rm s}y_{\rm e}$ are obtained by exchanging their subsystem bit configurations.

We sample the summation in equation~\eqref{eq:QCA_purity} by drawing two independent initial configurations $a,b$ from the diagonal distribution $p_0(a)=p(a,a)$. We evolve them to $x=f_t(a)$ and $y=f_t(b)$, exchange their subsystem bits, and evolve the resulting configurations backwards to $a'=f_t^{-1}(\widetilde{x})$ and $b'=f_t^{-1}(\widetilde{y})$. Then, assigning each pair the weight
\begin{align}
W(a,b)=\frac{p(a,a')}{p(a,a)}\frac{p(b,b')}{p(b,b)}
\label{eq:QCA_weight}
\end{align}
gives $\mathbb E_{a,b\sim p_0}[W(a,b)]=\Tr[\rho_{\rm s}(t)^2]$. 

At a finite time $t$, only a finite region of the initial state can influence the subsystem through the circuit gates. To identify this region, we start from the subsystem at time $t$ and trace the circuit backwards, layer by layer, including all input sites of every gate connected to the sites already retained. Because each layer contains finite-range gates, the number of sites in this region grows at most linearly with $t$. We can therefore perform the above sampling on the finite circuit formed by these gates, using the initial reduced density matrix on this region. Sites outside it cannot affect $\rho_{\rm s}(t)$, so this calculation gives the result in the thermodynamic limit $L\to\infty$ without finite-size saturation and provides a direct test of the predictions obtained from RP resonances.

For example, the initial states $\rho_{\rm TI}$ and $\rho_{\rm defect}$ used in the main text are product states built from $\rho_Z(m)=(I+mZ)/2$ and $I/2$. To sample $\rho_{\rm TI}$ in equation~\eqref{eq:rho_TI} of the main text, we independently draw the bit at each site $j=0\pmod4$ as $a_j=0$ with probability $(1+m)/2$ and $a_j=1$ with probability $(1-m)/2$. At all other sites, we draw $0$ and $1$ with equal probability. For $\rho_{\rm defect}$ in equation~\eqref{eq:rho_defect} of the main text, we use the same biased probabilities on the three defect sites $j\in\mathcal D$, and equal probabilities elsewhere. This procedure gives the Monte Carlo results in Fig.~\ref{fig:strong_QME}\textbf{c} of the main text.

\subsection{Block decomposition of the truncated propagator}
The permutation property in equation~\eqref{eq:QCA_map} also implies that the time evolution of a diagonal operator remains diagonal. Consequently, Pauli strings containing only $I$ and $Z$ form an invariant operator space, and their orthogonal complement is invariant as well. Therefore, the truncated propagator decomposes as 
\begin{align}
\mathcal E_{k,r}=\mathcal E^Z_{k,r}\oplus\widetilde{\mathcal E}_{k,r},
\label{eq:QCA_blocks}
\end{align}
where $\mathcal E^Z_{k,r}$ acts on the Fourier transforms of the Pauli strings in
\begin{align}
\mathscr P^Z_{r,s}
=\left\{\mathcal T^a\!\left(Z_1\prod_{j=2}^{r}P_j\right)
\,\middle|\,a=0,\ldots,s-1,\ P_j\in\{I_j,Z_j\}\right\}, 
\label{eq:QCA_Z_basis}
\end{align}
The complementary block $\widetilde{\mathcal E}_{k,r}$ acts on the Fourier transforms of Pauli strings containing at least one $X$ or $Y$ operator. The diagonal block $\mathcal E^Z_{k,r}$ has dimension $s2^{r-1}$, which is much smaller than $3s4^{r-1}$ for the full space. The complementary block can be further decomposed according to the parity of the number of $Y$ operators in a Pauli string. Indeed, $U$ is real in the computational basis, so a Pauli string satisfying $P^*=(-1)^{n_Y(P)}P$ evolves into a linear combination of Pauli strings with the same parity of $n_Y$, where $\bullet^*$ stands for the complex conjugate. Consequently, the full truncated propagator can be written as
\[
\mathcal E_{k,r}
=\mathcal E^Z_{k,r}
\oplus\widetilde{\mathcal E}^{\rm even}_{k,r}
\oplus\widetilde{\mathcal E}^{\rm odd}_{k,r}.
\]
Here $\widetilde{\mathcal E}^{\rm even}_{k,r}$ and $\widetilde{\mathcal E}^{\rm odd}_{k,r}$ act on the Fourier transforms of Pauli strings in the complementary space with an even and an odd number of $Y$ operators, respectively. For this QAC, we numerically find that the dominant RP mode belongs to the diagonal block $\mathcal E^Z_{k,r}$ at $k=0$.

\subsection{Cutoff convergence}
\label{sec:QCA_cutoff}
\begin{figure}[h]
\centering
\includegraphics[width=\textwidth]{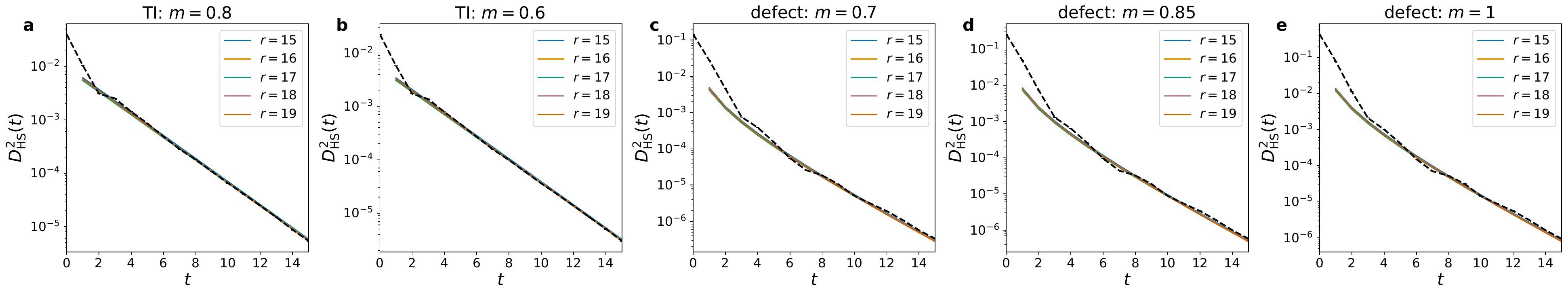}
\caption{\textbf{Cutoff convergence in the QAC.} Coloured solid curves show $D_{\rm HS}^2(t)$ predicted by the dominant RP resonance and its mode for $r=15,16,17,18,19$; black dashed curves show Monte Carlo results. For each initial state, we take $2^{27}$ samples in panels \textbf{a} and \textbf{b}, and $2^{33}$ samples in panels \textbf{c}--\textbf{e}. Panels \textbf{a} and \textbf{b} use $\rho_{\rm TI}$ in equation~\eqref{eq:rho_TI} of the main text with $m=0.8$ and $0.6$, respectively. Panels \textbf{c}--\textbf{e} use $\rho_{\rm defect}$ in equation~\eqref{eq:rho_defect} of the main text with $m=0.7,0.85,1$, respectively. The subsystem is $S=\{0,1,2,3\}$ and the defect sites are $\mathcal D=\{0,1,2\}$.}
\label{fig:QCA_cutoff}
\end{figure}

We also check the dependence of the QAC results in Fig.~\ref{fig:strong_QME}\textbf{c} of the main text on the operator-support cutoff $r$. Figure~\ref{fig:QCA_cutoff} compares the predictions obtained from the dominant RP resonance and its mode for $r=15,16,17,18,19$ with the Monte Carlo results for the initial states used in the main text. The curves for different $r$ are nearly indistinguishable over the times shown.

\section{Crossover of the relaxation law for spatially periodic initial states}\label{sec:QCA_crossover}
For an initial state invariant under translations by $sl$ sites, the overlap $\Tr_{}[\rho(0)V_{\alpha,k}^{\rm R}]$ selects discrete momenta, and the asymptotic relaxation is purely exponential for any finite $l$. On the other hand, in the limit $l\to\infty$, the initial state has no finite spatial period. In this limit, the relaxation acquires an algebraic correction, as discussed in Sec.~\ref{sec:Laplace}. Here, we investigate how these two relaxation laws are connected as the period $l$ is varied.

For simplicity, we consider a dominant RP resonance with a quadratic maximum of its magnitude at $k=0$ and the local expansion
\begin{align}
\lambda_{1,k}\simeq\lambda_{1,0}\exp\!\left(-\frac{\nu_{1,0}''}{2}k^2\right). 
\label{eq:QCA_crossover_band}
\end{align}
The spatial period restricts the momenta to $k_n=2\pi n/l$, taken in $[-\pi,\pi)$. In this case, according to equation~\eqref{eq:main_discrete} of the main text, the reduced density matrix takes the form
\begin{align}
\rho_{\rm s}(t)-\frac{I}{2^\ell}
\simeq \lambda_{1,0}^t
\sum_{k_n\in \Lambda_l}
c_{1,k_n} \varrho_{1,k_n}
e^{-\frac{t\nu_{1,0}''}{2}k_n^2}.
\label{eq:QCA_crossover_sum}
\end{align}
The sum in equation~\eqref{eq:QCA_crossover_sum} can be replaced by an integral when the momentum spacing $\Delta k=2\pi/l$ is much smaller than the Gaussian width $[t\nu_{1,0}''/2]^{-1/2}$. In this regime, many discrete momenta contribute, and $\sum_{k\in \Lambda_l}$ can be approximated by $l \int dk/(2\pi)$. Assuming that $c_{1,k}\varrho_{1,k}$ varies negligibly across this width, at times sufficiently late for the contribution from the dominant RP resonance and the quadratic expansion to apply, we obtain
\begin{align}
\rho_{\rm s}(t)-\frac{I}{2^\ell}
\simeq\lambda_{1,0}^t l
\int_{-\pi}^{\pi}\frac{dk}{2\pi} 
e^{-\frac{t\nu_{1,0}''}{2}k^2}c_{1,k} \varrho_{1,k}
\simeq\frac{\lambda_{1,0}^tlc_{1,0}\varrho_{1,0}}
{\sqrt{2\pi t\nu_{1,0}''}}.
\label{eq:QCA_crossover_integral}
\end{align}
Thus, even at finite $l$, the relaxation in this time window has the same $t^{-1/2}$ algebraic correction as in the infinite-period limit.

As time increases, the Gaussian narrows until its width becomes comparable to the momentum spacing. The approximation $\sum_{k\in \Lambda_l}\simeq l \int_{-\pi}^\pi \frac{dk}{2\pi}$ then breaks down. 
We thus expect that the crossover in the relaxation law occurs at $t\sim t_c$ such that 
\begin{align}
\frac{2\pi}{l}\sim\left(\frac{t\nu_{1,0}''}{2}\right)^{-1/2}
\quad\Longrightarrow\quad
t\sim t_c=\frac{l^2}{2\pi^2\nu_{1,0}''}.
\label{eq:QCA_crossover_time}
\end{align}
For $t\ll t_c$, equation~\eqref{eq:QCA_crossover_integral} gives the algebraically corrected relaxation. For $t\gg t_c$, the nonzero momenta are suppressed relative to $k=0$, and equation~\eqref{eq:QCA_crossover_sum} reduces to $\lambda_{1,0}^tc_{1,0}\varrho_{1,0}$. The relaxation therefore crosses over to a pure exponential at fixed finite period.

\begin{figure}[t]
\centering
\includegraphics[width=0.5\textwidth]{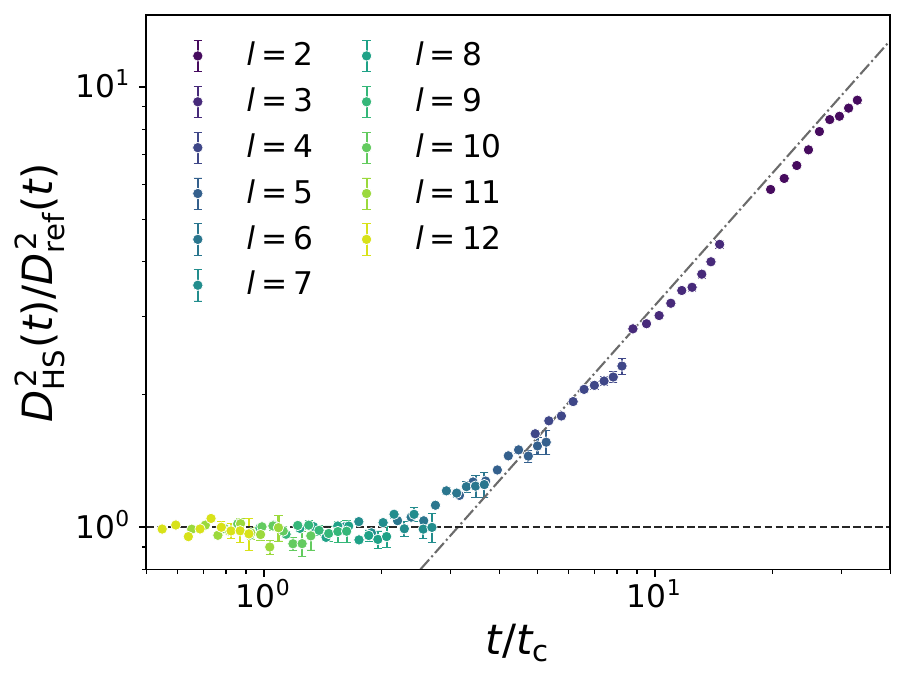}
\caption{\textbf{Crossover of the relaxation law for the spatially periodic initial state in equation~\eqref{eq:QCA_periodic_initial}}. Symbols show the Monte Carlo squared distance divided by $D_{\rm ref}^2(t)$ in equation~\eqref{eq:QCA_crossover_reference}, plotted as a function of $t/t_c$. The dash-dotted line represents $t/(\pi t_c)$. Error bars indicate standard errors of the normalised Monte Carlo results. The subsystem is $S=\{0,1,2,3\}$, the plotted times are $t=12,\ldots,20$, and the sample count is $2^{32}$ per initial state. The reference curve and $t_c$ are evaluated using RP resonances at $r=13$.}
\label{fig:QCA_crossover}
\end{figure}

To test these predictions numerically, we consider the dynamics of the QAC introduced in Sec.~\ref{sec:QCA} with the following initial state
\begin{align}
\rho_l(0)=\bigotimes_{j\in\mathbb Z}\rho_Z(m_j),\qquad
m_j=\begin{cases}
1&j=0\pmod{sl},\\
0&\text{otherwise},
\end{cases}
\label{eq:QCA_periodic_initial}
\end{align}
where $s=4$. We write $D_{\rm HS}^2(t)$ for the squared Hilbert--Schmidt distance from equilibrium for this state. At finite $l$, equations~\eqref{eq:QCA_crossover_sum} and~\eqref{eq:QCA_crossover_integral} yield
\begin{align}
D_{\rm HS}^2(t)\simeq
\begin{dcases}
D_{\rm ref}^2(t)&t\ll t_c,\\
|\lambda_{1,0}|^{2t}|c_{1,0}|^2
\Tr[\varrho_{1,0}^{\dagger}\varrho_{1,0}],&t\gg t_c, 
\end{dcases}
\label{eq:QCA_crossover_finite}
\end{align}
with 
\begin{align}
D_{\rm ref}^2(t):=
\frac{|\lambda_{1,0}|^{2t}l^2|c_{1,0}|^2}{2\pi t\nu_{1,0}''}
\Tr[\varrho_{1,0}^{\dagger}\varrho_{1,0}].
\label{eq:QCA_crossover_reference}
\end{align}
Normalising by this reference curve at all plotted times, we expect
\begin{align}
\frac{D_{\rm HS}^2(t)}{D_{\rm ref}^2(t)}
\simeq\begin{cases}
1,&t\ll t_c,\\[2pt]
\displaystyle\frac{2\pi t\nu_{1,0}''}{l^2}
=\displaystyle\frac{t}{\pi t_c},&t\gg t_c.
\end{cases}
\label{eq:QCA_crossover_ratio}
\end{align}
Figure~\ref{fig:QCA_crossover} tests equation~\eqref{eq:QCA_crossover_ratio} using Monte Carlo simulations. At small $t/t_c$, the ratio remains close to unity, consistent with the $t^{-1}$ correction to the squared Hilbert--Schmidt distance at intermediate times. At larger $t/t_c$, it grows approximately as $t/(\pi t_c)$, confirming the expected crossover to purely exponential relaxation. 

\section{Relaxation law for random product states}\label{sec:random_product}
Complete translation-symmetry breaking can also be realised by random product states such as 
\begin{align}
\rho(0)=\bigotimes_{j\in\mathbb Z}\rho_Z(m\omega_j),\qquad
\mathbb P(\omega_j=+1)=\mathbb P(\omega_j=-1)=\frac12,
\label{eq:random_initial}
\end{align}
However, the steepest-descent analysis in Sec.~\ref{sec:Laplace} cannot be applied directly to such initial states, because complete translation-symmetry breaking alone does not ensure that the overlap $\Tr_{}[\rho(0)V_{\alpha,k}^{\rm R}]$ is a smooth function of $k$. In fact, if we rewrite the overlap for the initial state~\eqref{eq:random_initial} as 
\begin{align}
\Tr[\rho(0)V_{\alpha,k}^{\rm R}]
=\sum_{P\in\mathscr P^Z_{r,s}}v_{\alpha,P_k}^{\rm R}
\sum_{n\in\mathbb Z}e^{-\im kn}a_{P,n},\quad a_{P,n}=\Tr[\rho(0)\mathcal T^{sn}P]
=m^{N_Z(P)}\prod_{j:\,P_j=Z_j}\omega_{j+sn},
\label{eq:random_overlap}
\end{align}
with $N_Z(P)$ being the number of $Z$ operators in $P$, one finds that $|a_{P,n}|=m^{N_Z(P)}$ does not tend to zero as $|n|\to\infty$. By the Riemann--Lebesgue lemma, each Fourier series $\sum_n e^{-\im kn}a_{P,n}$ is therefore not an integrable function of $k$. In general, neither is their combination $\Tr_{}[\rho(0)V_{\alpha,k}^{\rm R}]$, so the smoothness of $\Tr_{}[\rho(0)V_{\alpha,k}^{\rm R}]$ used in Sec.~\ref{sec:Laplace} is not guaranteed. 

To investigate relaxation in this case, we consider $\mathbb E[D_{\rm HS}^2(t)]$ for the initial state in equation~\eqref{eq:random_initial}, where $\mathbb E$ denotes the average over the independent random signs $\omega_j$. Using equation~\eqref{eq:integral}, $\mathbb E[D_{\rm HS}^2(t)]$ can be written as 
\begin{align}
\mathbb E[D_{\rm HS}^2(t)]
\simeq\int_{-\pi}^{\pi}\frac{dk\,dq}{(2\pi)^2}
\lambda_{1,k}^t(\lambda_{1,q}^*)^t
\Tr[\varrho_{1,k}\varrho_{1,q}^{\dagger}]
\mathbb E\!\left[
\Tr[\rho(0)V_{1,k}^{\rm R}]
\Tr[\rho(0)V_{1,q}^{\rm R}]^*\right].
\label{eq:random_distance_double_integral}
\end{align}
Substituting equation~\eqref{eq:random_overlap} into the average in this expression gives
\begin{align}
\mathbb E\!\left[
\Tr[\rho(0)V_{1,k}^{\rm R}]
\Tr[\rho(0)V_{1,q}^{\rm R}]^*\right]
=\sum_{n,n'\in\mathbb Z}e^{-\im kn+\im qn'}
\sum_{P,Q\in\mathscr P^Z_{r,s}}
 v_{1,P_k}^{\rm R}(v_{1,Q_q}^{\rm R})^*
\mathbb E[a_{P,n}a_{Q,n'}].
\label{eq:random_overlap_expansion}
\end{align}
We evaluate the remaining average using the independence and zero mean of the signs. Unless $\mathcal T^{sn}P$ and $\mathcal T^{sn'}Q$ contain $Z$ on exactly the same sites, at least one sign appears only once in their product, and the average vanishes. Since the Pauli strings are anchored by their leftmost nonidentity site, the surviving terms have $P=Q$ and $n=n'$, for which $\mathbb E[a_{P,n}a_{Q,n'}]=m^{2N_Z(P)}$. Equation~\eqref{eq:random_overlap_expansion} consequently becomes
\begin{align}
\mathbb E\!\left[
\Tr[\rho(0)V_{1,k}^{\rm R}]
\Tr[\rho(0)V_{1,q}^{\rm R}]^*\right]
=2\pi\delta(k-q)b_k,\quad 
b_k=\sum_{P\in\mathscr P^Z_{r,s}}
|v_{1,P_k}^{\rm R}|^2m^{2N_Z(P)}. 
\label{eq:random_overlap_covariance}
\end{align}
Substituting equation~\eqref{eq:random_overlap_covariance} into equation~\eqref{eq:random_distance_double_integral} and performing the $q$ integral, we obtain
\begin{align}
\mathbb E[D_{\rm HS}^2(t)]
\simeq\int_{-\pi}^{\pi}\frac{dk}{2\pi}
|\lambda_{1,k}|^{2t}b_k\Tr[\varrho_{1,k}^{\dagger}\varrho_{1,k}].
\label{eq:random_distance_integral}
\end{align}
We can now apply steepest descent to this single integral, since $b_k$ and $\Tr_{}[\varrho_{1,k}^{\dagger}\varrho_{1,k}]$ are smooth. It results in
\begin{align}
\mathbb E[D_{\rm HS}^2(t)]\simeq
\frac{|\lambda_{1,0}|^{2t}b_0\Tr[\varrho_{1,0}^{\dagger}\varrho_{1,0}]}
{\sqrt{4\pi t \nu_{1,0}''}}.
\label{eq:random_distance_asymptotic}
\end{align}
The algebraic correction to the mean squared distance is therefore $t^{-1/2}$, rather than $t^{-1}$ predicted by equation~\eqref{eq:asymptotic} for initial states with smooth overlap. 

\begin{figure}
\centering
\includegraphics[width=0.5\textwidth]{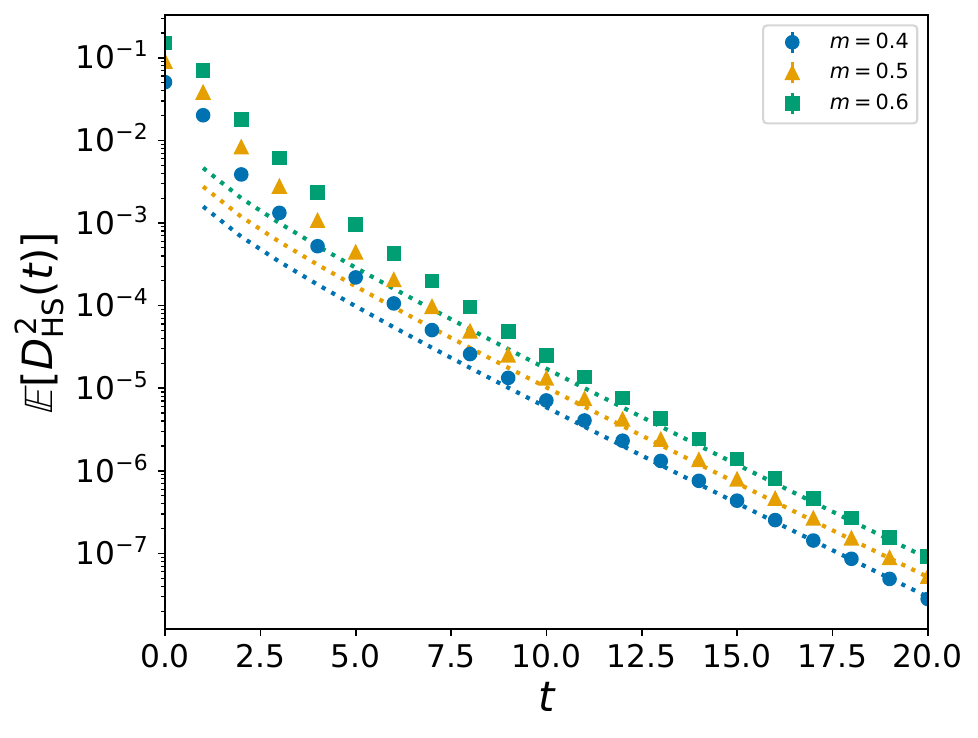}
\caption{\textbf{Relaxation from random product states in equation~\eqref{eq:random_initial}}. Symbols show the mean squared Hilbert--Schmidt distance obtained by Monte Carlo simulation and averaging over the random signs. Dotted curves show the prediction in equation~\eqref{eq:random_distance_asymptotic}, obtained from the dominant RP resonance and its mode calculated with $r=13$. Statistical error bars are smaller than the symbols. The subsystem size is $\ell=4$ and the average uses $2^{10}$ initial-state realisations with $2^{24}$ Monte Carlo samples for each realisation.}
\label{fig:QCA_random}
\end{figure}

To test the prediction in equation~\eqref{eq:random_distance_asymptotic}, we use the QAC introduced in Sec.~\ref{sec:QCA}. For each realisation in equation~\eqref{eq:random_initial}, we compute $D_{\rm HS}^2(t)$ by Monte Carlo sampling and then average over realisations. Figure~\ref{fig:QCA_random} shows the results together with equation~\eqref{eq:random_distance_asymptotic}, evaluated from the RP resonance and its eigenoperators. The numerical curves approach the predicted asymptotic form at late times, supporting the $t^{-1/2}$ correction to the mean squared distance, as expected.

% References are provided directly with bibitem entries.

\end{document}